\documentclass[journal,compsoc]{IEEEtran}
\usepackage{amsmath,amsfonts}
\usepackage{algorithmic}
\usepackage{algorithm}
\usepackage{array}
\usepackage[caption=false,font=normalsize,labelfont=sf,textfont=sf]{subfig}
\usepackage{textcomp}
\usepackage{stfloats}
\usepackage{url}
\usepackage{verbatim}
\usepackage{graphicx}
\usepackage{cite}
\usepackage{xcolor}
\usepackage{tikz}

\usepackage{cleveref}
\crefname{figure}{Fig.}{Figures}
\Crefname{figure}{Fig.}{Figures}

\begin{document}

\newcommand{\note}[1]{{\color{black}#1}}

\newcommand{\sysname}[1]{\textit{AgentLens}}

\newcommand{\mybox}[1]{\tikz[baseline=(MeNode.base)]{\node[circle, fill=gray!20, minimum size=0.1em, text centered,inner sep=0pt, font=\fontsize{8pt}{1em}\selectfont](MeNode){\strut#1}}}

\title{AgentLens: Visual Analysis for Agent Behaviors in LLM-based Autonomous Systems}

\author{Jiaying Lu, Bo Pan, Jieyi Chen, Yingchaojie Feng, Jingyuan Hu, Yuchen Peng, Wei Chen
}

\markboth{Journal of \LaTeX\ Class Files,~Vol.~14, No.~8, August~2021}%
{Shell \MakeLowercase{\textit{et al.}}: A Sample Article Using IEEEtran.cls for IEEE Journals}

\IEEEpubid{0000--0000/00\$00.00~\copyright~2021 IEEE}

\maketitle

\begin{abstract}
Recently, Large Language Model based Autonomous system (LLMAS) has gained great popularity for its potential to simulate complicated behaviors of human societies. One of its main challenges is to present and analyze the dynamic events evolution of LLMAS. In this work, we present a visualization approach to explore detailed statuses and agents’ behavior within LLMAS. We propose a general pipeline that establishes a behavior structure from raw LLMAS execution events, leverages a behavior summarization algorithm to construct a hierarchical summary of the entire structure in terms of time sequence, and a cause trace method to mine the causal relationship between agent behaviors. We then develop \sysname{}, a visual analysis system that leverages a hierarchical temporal visualization for illustrating the evolution of LLMAS, and supports users to interactively investigate details and causes of agents' behaviors. Two usage scenarios and a user study demonstrate the effectiveness and usability of our \sysname{}.

\end{abstract}


\begin{figure*}
    \centering
    \includegraphics[width=1\linewidth]{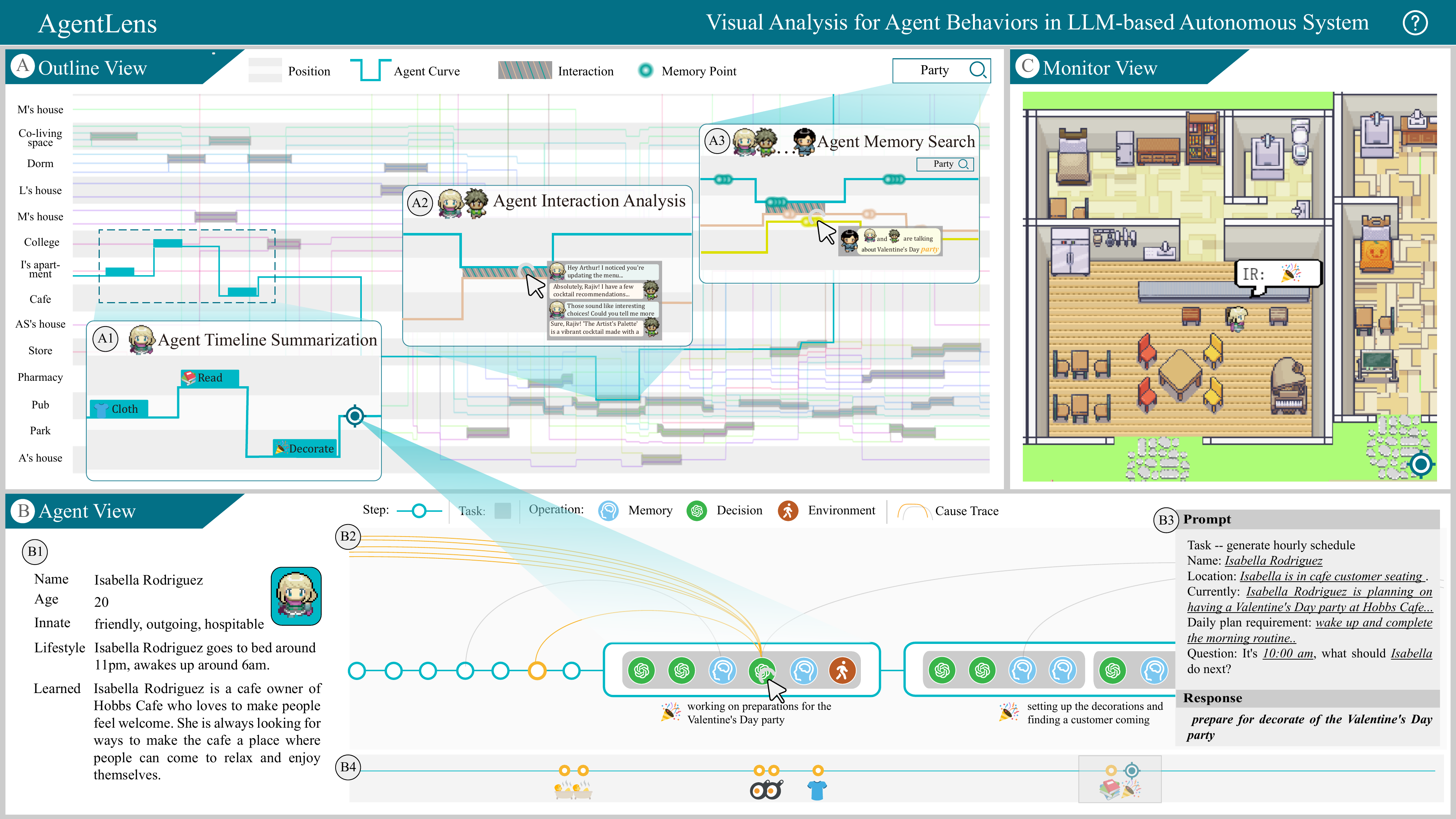}
    \caption{The user interface of \sysname{} comprises three views. The \textit{Outline View} (A) displays the trajectory of each agent using different colored curves, enabling users to identify significant patterns or event summarization during the evolution of LLMAS. By clicking on a time step in each curve, users can further investigate it in the \textit{Agent View} (B). It allows users to progressively reveal agent event information and trace the cause of specific agent behavior. The \textit{Monitor View} (C) automatically adjusts the \note{graphical representation} of LLMAS based on the user's current point of interest.}
    \label{fig:teaser}
\end{figure*}

\section{Introduction}
\IEEEPARstart{A}{utonomous} agents, as computational entities that possess a certain degree of autonomy\cite{agha1986model,green1997software}, are seen as a promising pathway toward achieving artificial general intelligence (AGI)\cite{wooldridge1995intelligent,hutter2004universal}.
In recent years, owing to the breakthroughs in natural language processing\cite{ouyang2022training,openai2023gpt4,wei2022emergent} achieved by Large Language Models (LLM), the LLM-based autonomous agent has gained widespread adoption in both academia and industry\cite{wang2023survey, xi2023rise}. 
\note{Built upon LLM-based agents, LLM-based autonomous systems (LLMAS) deploy multiple agents within a shared environment, enabling them to display behavior and social patterns akin to humans. 
This collective intelligence fosters emergent social dynamics, such as the formation of new relationships, diffusion of information, and the rise of coordination among agents\cite{park2023generative}.}
Consequently, LLMAS exhibits significant potential in society simulation\cite{park2023generative,jinxin2023cgmi}, software engineering\cite{qian2023communicative,hong2023metagpt}, and scientific research\cite{boiko2023emergen}.

However, monitoring and analyzing the dynamic evolution of LLMAS, \note{including agents in LLMAS and event sequences undertaken by them,} can be challenging due to the tremendous information generated during the system evolution and the inherent unpredictability of LLMs. The most straightforward approach for analyzing LLMAS is to inject logging code into LLMAS to trace agent events of interest and check the raw output logs in text format\cite{AutoGPT}. However, this approach requires expertise with specific LLMAS and is unintuitive for general users. To address this, many LLMAS projects provide a \note{graphical representation} of the simulation process\cite{xi2023rise}, which is typically re-playable 2D \cite{lin2023agentsims,chen2023agentverse,park2023generative,zhang2023proagent} or 3D video\cite{zhang2023building,zhu2023ghost,hafner2023mastering,mirchev2021variational}. By transforming a fixed sequence of intermediate simulation \note{events} into expressive visual recordings, users can digest that information more efficiently and intuitively. However, a re-playable recording with a fixed level of abstraction limits the flexibility of analysis for LLMAS. Even for a specific LLMAS and a fixed usage scenario, a user's short-term analysis target will change frequently during the analysis process. As the users' analysis target varies, the type, quantity, and granularity of agent events to be visualized also need to change. Moreover, analyzing the agent's behavior at a specific time point requires users to switch the recording back and forth to trace the cause and consequence of this behavior, which is tedious and unreliable.

This work thus presents a visualization approach to assist users in efficiently analyzing \note{the evolving status and complex behaviors of agents within an LLMAS}. \note{To mitigate cognitive overload due to the profusion of data produced throughout the evolution of LLMAS, and to enhance adaptability for subsequent analytical processes, we introduce a general pipeline, which establishes a hierarchical behavior structure of agent entities and raw event sequences within the LLMAS operational records. The formulation of the structure is based on our survey of prevalent architectures within extant LLMAS, coupled with a design study that engaged 4 LLMAS developers and 4 layman users.} \note{We design an LLM-based algorithm for summarizing agent behavior that furnishes a hierarchical depiction of sequences of agent events. Additionally, we employ a cause trace method to unearth the causal linkages among disparate agent events.} Based on the extracted hierarchical structure, we then develop \sysname{}, a visual analysis system designed to facilitate interactive analysis and exploration of agent behaviors in LLMAS.

\note{ \sysname{} provides a multi-faceted perspective for LLMAS through its three distinct but interrelated views, each offering a different level of abstraction}. The \textit{Outline View} (\cref{fig:teaser}, \mybox{A}) illustrates the spatiotemporal trajectory of each agent with curves of different colors, \note{aiding users in identifying notable agents or their intriguing behaviors throughout the evolution of LLMAS.} Users can quickly scan agent behaviors at different granularity (\cref{fig:teaser}, \mybox{$A_1$}), identify agent interaction of interest (\cref{fig:teaser}, \mybox{$A_2$}), perform topic search (\cref{fig:teaser}, \mybox{$A_3$}), and click any time point on an agent curve to further investigate it in the \textit{Agent View} (\cref{fig:teaser}, \mybox{B}). The \textit{Agent View} allows users to progressively reveal agent event information on demand and trace the cause of certain agent behavior. The \textit{Monitor View} (\cref{fig:teaser}, \mybox{C}) automatically adjusts the \note{graphical representation} of LLMAS for users based on their current point of interest in the \textit{Outline View} or the \textit{Agent View}.  
To evaluate the performance of \sysname{}, we present two cases and conduct a user study with 14 participants to gather their feedback. The results indicate that \sysname{} is capable of assisting users in the LLMAS evolution analysis and agent behaviors investigation.

The main contributions of our work are as follows:

\begin{itemize}
    
    \item To the best of our knowledge, our work is the first visual analysis system that enables analysis and explorations of agent behaviors within LLMAS.
    \item We propose a general pipeline that establishes a hierarchical behavior structure from raw LLMAS execution events to facilitate downstream analysis.
    
    \item We conduct two cases and a user study to demonstrate the capabilities of our system. The evaluation results confirm the usefulness and effectiveness of the behavior structure and\sysname{}.
\end{itemize}

\section{Related Work}

\subsection{LLM-based Autonomous Agents}

Franklin \textit{et al.}\cite{franklin1996agent} defined the agent as an entity situated in the environment that senses the environment and acts on it over time, in pursuit of its own agenda and so as to affect what it senses in the future. 
Possessing the ability to perform intelligent operations without human intervention, the autonomous agent remains a steadfast goal in artificial intelligence research\cite{wooldridge1995intelligent,bubeck2023sparks}.

The progression of LLMs \cite{touvron2023llama,openai2023gpt4} has underscored exceptional proficiency in areas of comprehension, reasoning, and language generation\cite{brown2020language}, which kindled optimism for continued advancements in the realm of autonomous agents.
With the advent of LLMs, the study of LLM-based autonomous agents began to thrive. This includes enhancing agents' self-reflective capabilities \cite{yao2023react,shinn2023reflexion}, implementing superior task decomposition strategies \cite{yao2023tree}, and endowing the ability to utilize and create tools\cite{schick2023toolformer,qin2023tool,qin2023toolllm,qian2023creator}.
There is also a vibrant development of applications of LLM-based agents in the open source community \cite{Langchain,BabyAGI,AutoGPT}.

Recently, researchers have found that LLM-based agents can address a wider range of tasks through collaboration or competition.
Camel\cite{li2023camel} presented a framework that emphasizes the autonomous interaction between communicative agents. It is capable of creating varied, detailed instructions across numerous tasks, thereby providing a platform for these agents to demonstrate their cognitive operations.   
Talebirad \textit{et al.}\cite{talebirad2023multi} introduced a comprehensive framework for multi-agent collaboration based on LLMs.
ProAgent\cite{zhang2023proagent} exhibited the distinctive ability for agents to foresee the upcoming decisions of collaborators and adjust their behaviors, enabling them to excel in cooperative reasoning tasks.
Multi-Agent Debate (MAD)\cite{liang2023encouraging} introduced an approach in which several agents present their arguments collaboratively while a judge guides the discourse, enhancing agents' divergent thinking for deep-reflective tasks.

However, as the number and the intricacy of agents increase, the complexity of analyzing their behaviors escalates rapidly.
While past works have focused on elevating the capabilities of LLM-based agents in emulating human-like behaviors, they often overlooked how to effectively analyze agent behaviors. In this work, we identify this research gap and present a visualization approach for analyzing agent behaviors in LLM-based multi-agent systems.

\subsection{LLM-based Autonomous System}

By incorporating numerous LLM-based agents into a cohesive environment, the LLMAS is capable of handling diverse complex scenarios.
For example, WebAgent\cite{nakano2022webgpt} demonstrated the possibility of building agents that can complete the tasks on real websites following natural language instructions. ChatDev\cite{qian2023communicative} and MetaGPT\cite{hong2023metagpt} experimented with software development in multi-agent communication settings. Zhang \textit{et al.}\cite{zhang2023building} built embodied agents to cooperate effectively with humans.
Park \textit{et al.}\cite{park2023generative} situates generative agents with unique characteristics in a societal context, in order to mimic human social behaviors.

Several task-independent frameworks designed for diverse usages have received considerable attention within the community. 
AgentVerse\cite{chen2023agentverse} dynamically assembled multi-agent teams tailored to task complexities, outperforming individual agents with adaptable team structures. AgentSims\cite{lin2023agentsims} offered a real-time evaluation platform for LLM-based agents, enabling adaptable configurations to facilitate the performance evaluation of different modules. AutoGen\cite{wu2023autogen} fostered conversations among multiple agents and organized individual insights in a general manner, offering an interconnected manner to coordinate multiple agents within the LLMAS. 
\note{MetaGPT\cite{hong2023metagpt} injects effective human workflows into multi-agent collaboration by encoding Standardized Operational Procedures (SOP) into prompts, underscoring the potential of incorporating human domain expertise into LLMAS.}
CGMI\cite{jinxin2023cgmi} replicated human interactions and imitated human routines in real-world scenarios, which enhances the realism of more humanized simulation of complex social scenarios.

Previous LLMAS research has primarily focused on constructing more universal frameworks or designing for specific domains, yet there has been a noticeable lack of emphasis on the analysis methods of parallel behaviors among agents within LLMAS.
Contemporary LLMAS predominantly depend on conventional methods for surveillance and analysis. MetaGPT\cite{hong2023metagpt} utilizes log outputs for record maintenance, while Park et al.\cite{park2023generative} adopts panoramic videos for observation, providing detailed maps with agent avatars to denote their locations and behaviors. 
Distinct from preceding efforts, our work offers an interactive visual system that hierarchically organizes events, facilitating users in quickly grasping the happenings within LLMAS.

\subsection{Event Sequence Visualization}
Data featuring time-based event sequences is widespread and can be found in various sectors, including healthcare records\cite{guo2020comparative,jin2020carepre,nielsen2009abyss}, career design\cite{guo2018eventthread,guo2018visual} and social interactions\cite{perer2014frequence,han2015visual,cao2015episogram}. In these fields, distinct types of time-stamped events are sequentially organized, each relevant to a particular subject or entity. While earlier methods\cite{fischer2012vistracer,8265216} have been geared toward simpler, low-dimensional data, the data sets encountered in real-world scenarios frequently display a higher level of complexity, calling for more comprehensive analytical ideas and methods.

\note{A substantial number of research on event sequence visualization is notably correlated with fields where there is a prevalent demand for event information condensation, such as in the realm of social media data\cite{wu2016survey}, the sphere of smart manufacturing \cite{zhou2019survey}, and the study of anomalous user behaviors\cite{shi2019visual}. 
Guo \textit{et al.}\cite{guo2020survey}  proposed an organizational framework for event sequences to summarize the common goal of different properties with great heterogeneity.
EventThread\cite{guo2018eventthread} focuses on visualization and cluster analysis, providing an interactive interface for browsing and summarizing event sequence data.}
\note{
Building on past frameworks of condensing events and visualizing them, we focused on the behavioral patterns of LLM agents and proposed an LLM-driven approach to handle non-structured natural language-based event sequences.}

Event sequence visualization has highly relevant applications in the realm of collective behavior analysis, which aligns closely with the focus of our research, both referring to activities conducted by a temporary and unstructured group of people \cite{yuan2014visualization,wu2014opinionflow,han2015visual}. In the field of social media, collective actions emerge from the collaborative efforts of users engaged in disseminating information and navigating through virtual spaces. A variety of sophisticated visual analytics methodologies have been introduced to scrutinize these group dynamics. R-map\cite{chen2019r}, Socialwave\cite{sun2017socialwave}, FluxFlow\cite{zhao2014fluxflow} and Google+ ripples\cite{viegas2013google+} are specifically tailored to examine the mechanics of information propagation, while Maqui\cite{law2018maqui} and Frequence\cite{perer2014frequence} offers insights into the complexities of human mobility within this context.

While existing research has made significant contributions to the field, there's a growing need to address the increasingly complex behaviors and interactions that call for the advancement of autonomous systems. Our work introduces event sequence visualization as an integral tool for the analysis and exploration of LLMAS.

\section{Overview}
\label{sec:Overview}
\subsection{Common Architecture of LLMAS}
\label{sec:comm_arch}
\begin{figure}[t!]
    \centering
    \includegraphics[width=1\linewidth]{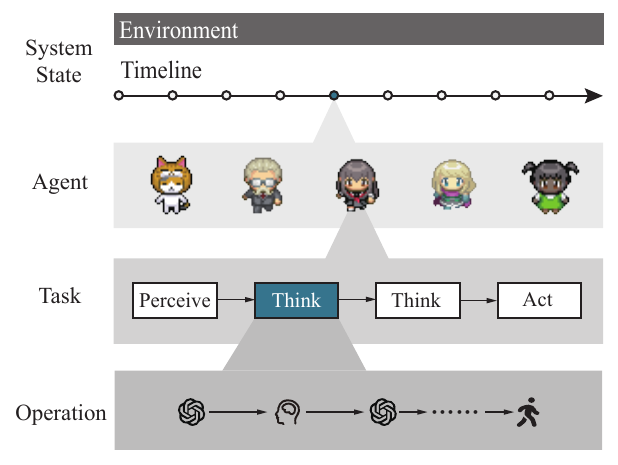}
    \caption{The common architecture abstracted from existing LLMAS consists of four layers: system states, agents, tasks, and operations.
    }
    \label{fig:structure}
\end{figure}
To ensure maximum compatibility with various LLMAS, we survey LLMAS-related papers
\cite{dambekodi2020playing,hausknecht2020interactive,liu2023training,driess2023palm,paul2022avlen,park2022social,gao2023s,williams2023epidemic,yao2023react,shinn2023reflexion} as well as some projects\cite{ogara2023hoodwinked,fan2022minedojo,wang2023voyager,lynch2023interactive,surís2023vipergpt,nottingham2023embodied,cui2023chatlaw,huang2022inner,shen2023hugginggpt,liang2023unleashing} with high stars in open source communities published before August 31, 2023.
We analyzed their system architectures and components, based on which we abstract a common architecture (as shown in \cref{fig:structure}) for LLMAS.
The \textbf{system state} in LLMAS provides the environmental information at any time point. 
At each time point, each \textbf{agent} executes its own \textbf{task}, which consists of several atomic \textbf{operations}.
\note{A raw event is generated whenever an operation is executed by an agent, thereby advancing the evolution of LLMAS.}

\textbf{System State}
provides a comprehensive understanding of the environment. By acquiring the environmental information from the system state, agents can comprehend the current context and conditions. For example, the system state can inform agents about object locations and environmental properties, which significantly impact their decision-making and planning processes.
In addition, the system state governs the timelines of each agent, ensuring events by different agents are temporally aligned.

\textbf{Agents}
are autonomous entities with cognitive abilities and action capability. 
By performing various types of tasks, agents can interact with the environment and gradually change the system state to achieve their goals. Additionally, agents can communicate and collaborate with each other. They can share their knowledge and exchange messages to accomplish more complex duties.

\textbf{Tasks}
are typically customized for the usage scenario of LLMAS. 
A sequence of operations with a common goal can be grouped as a task.
Extending prior research that has focused on different scenarios for agents, we classify tasks into three categories: Perceive, Think, and Act. 
In \textit{Perceive} tasks, the agent obtains perception of the external system. Such perception includes sensing the environment (virtual, real, or external resources), as well as perceiving other agents. 
In \textit{Think} tasks, the agent engages in decision-making, reasoning, planning, and other behaviors based on external perception and its own memory. 
In \textit{Act} tasks, the agent interacts with the external system by providing outputs, including text outputs, virtual actions, or specific invocations such as tool usage.

\textbf{Operations} are the basic units for Tasks.
Operations can be classified based on their target, including Environmental Operations, Memory Operations, and Decision Operations.
\textit{Environment Operation}s execute interactions toward the external system, including other agents and the environment defined by LLMAS. \textit{Memory Operation}s involve storing and updating the memory of an agent. \textit{Decision Operation}s are for decision-making and action planning, where LLM-based agents typically utilize LLMs for decision operations.

\subsection{Design Requirement}

Our study focuses on users involved in analyzing, exploring, and monitoring LLMAS.  Our primary goal is to create a system that enhances users' comprehension of LLMAS.
We recruited 4 developers highly familiar with LLMAS and 4 users who have a basic understanding of LLMAS and have previously utilized such systems.

To identify the design requirements, We asked participants to explore the behaviors of agents in Reverie\footnote{https://reverie.herokuapp.com/arXiv\_Demo/\#}, a typical autonomous system consisting of 25 LLM-based agents. 
We requested participants to actively explore and delve into the identification of agent behaviors that intrigued them, as well as to investigate the underlying causes or consequences. To facilitate this, we encouraged participants to ``think aloud'', articulating the information they sought and the type of assistance they desired throughout the process.
We then conducted the first interview with them to collect their feedback on the whole exploration process. At the same time, we maintain regular contact with them to keep them updated on the design requirements. Based on their feedback, and combined with the survey on existing LLMAS work in \cref{sec:comm_arch}, the following 4 design requirements can be summarized.

\textbf{R1. Provide suitable generality of information for different analysis targets.} During the evolution of LLMAS, a significant volume of information is continuously generated, which is overwhelming for users to comprehend. While the current 2D graphical interface of Reverie provides a fixed visual abstraction, many users express their desire to change the generality of presented information to better match their current analysis target. For instance, users want to scan summarized agent traces across a large time scale when they analyze the long-term relationship among several agents, while they prefer a detailed presentation of an agent's operations when they analyze how the agent performs a certain task. Therefore, the system should provide users with flexible levels of abstraction for the generated information of LLMAS, and allows users to reveal details according to their analysis target. 

\textbf{R2. Present agents' transition of physical location and thought content.}
The physical and mental changes of agents play a vital role in driving and reflecting the evolution of the entire LLMAS. Nevertheless, currently, users can only stare at the re-playable recording to see if there is a location transition of the agent and check the raw execution log to find when the agent starts to think about a certain idea, which is inefficient and error-prone. Therefore, the system should provide visual emphasis on agents' transition of location and highlight the time points the agent starts to think about a topic the user wishes to explore. 

\textbf{R3. Underscore possible causes of agent behaviors.}
When users become interested in a certain behavior of the agent, they usually want to investigate the cause or consequence of this behavior. However, an agent’s behavior can be influenced not only by its current perception and thoughts but also by the memory of its past behavior. It is tedious and unreliable for users to switch the replayable recording back and forth to locate the cause of the behaviors of certain agents. Therefore, the system should provide a mechanism to mine the possible causes of an agent's behaviors and highlight them for users' investigation.

\textbf{R4. Explicate the context of LLM invocation.}
LLM plays a crucial role as the core of the LLMAS, which is frequently invocated to make cognitive decisions for agents. To inform the background for making a certain decision, the preceding contextual information is organized in a specific manner with a customized template and then sent as a prompt to the LLM. Therefore, to help users understand how and why a decision is made by an agent, the system should present the decisions made by LLM and explicate the context of its invocation. Moreover, it is desirable to provide visual enhancement to help users trace how the context information is collected from previous agent behaviors.

\subsection{Approach Overview}

\begin{figure}[!h]
    \centering
    \includegraphics[width=1\linewidth]{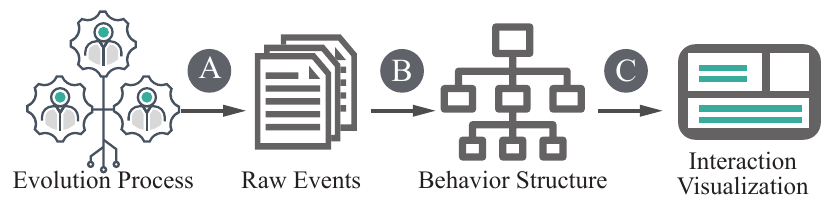}
    \caption{The workflow of our approach consists of three major steps. (A) Collect raw execution log of events from the LLMAS evolution process. (B) Establish a behavior structure with hierarchical summarization and a cause trace method. (C) Provide an interactive user interface for visual exploration and analysis.}
    \label{fig:overview}
\end{figure}

In alignment with the aforementioned design requirements, we designed \sysname{}, a proof-of-concept system dedicated to visualizing agent behaviors during the LLMAS evolution. The workflow of our approach is depicted in \cref{fig:overview}. 
\note{Users can utilize logging codes to log their LLMAS evolution process and capture raw events executed by agents. Based on these raw events, we establish a hierarchical structure to summarize agent behaviors in different granularity and trace possible causal relationships among their behaviors (\cref{sec:Structure Extraction}).} A user interface and a series of interactions are provided to support interactive exploration and analysis of the agent behaviors in LLMAS (\cref{sec:interface}).

\note{
\section{Behavior Structure Establishment}
\label{sec:Structure Extraction}
\begin{figure}[!h]
    \centering
    \includegraphics[width=1\linewidth]{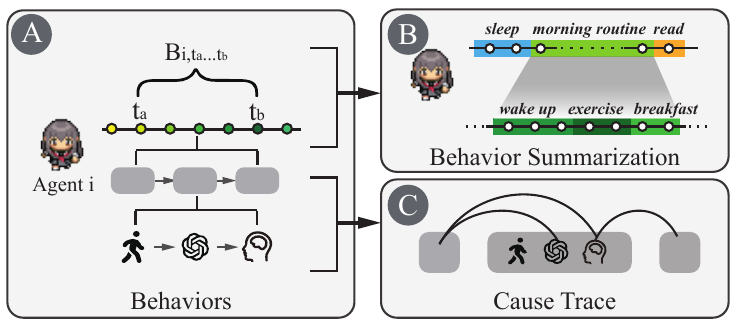}
    \caption{\note{The behavior structure is established through a three-step pipeline:  (A) We organize raw events into behaviors, (B) summarize and segment behaviors for an agent, and (C) trace causal relationships among behaviors.}} 
    \label{fig:approach}
\end{figure}

In this section, we introduce a pipeline designed to establish the hierarchical behavior structure from raw events generated during the evolution of LLMAS. It facilitates the generation of structured data for visualization, achieved via summarization and causal analysis of agent behaviors. As shown in \cref{fig:approach}, the pipeline consists of three steps: (A) processing the raw events and organizing them into behaviors based on the common architecture shown in \cref{fig:structure} (\textbf{R1}), (B) summarizing these behaviors and segmenting them in accordance with their semantic implications. (\textbf{R1}, \textbf{R2}), and (C) tracing the cause between these behaviors by analyzing the correlations among original events (\textbf{R3}, \textbf{R4}).

}

\note{
\subsection{Behavior Definition}
\label{sec:Behavior Definition}

During the evolution of LLMAS, multiple raw events are generated, creating large, often chaotic, and obscure text logs with the scaling of agent populations.
To streamline downstream analysis and visualization efforts, we defined agent behaviors as structured representations that encapsulate the sequence of raw events (\textbf{R1}).

Drawing upon the system state adopted by most LLMAS architectures, we denote the timeline $T$ to represent the states and events of agents at various time points within environments.
For each time point $t$ on the timeline, we can define the tuple $T_t$ as follows: 

\begin{equation}
    T_t = \langle e_{t-1}, \bigcup a_{t-1}[i], \bigcup s_t[i] \rangle
\end{equation}
where $e_{t-1}$ denotes the environment state at the previous point $t-1$ before time $t$. $a_{t-1}[i]$ represents the agent state of the $i$-th agent at $t-1$, encompassing its position within the environment as well as individual status indicators such as hunger levels, mood values, etc. In various LLMAS, $a_t$ encompasses a diverse array of attributes. $s_t[i]$ denotes the set of indivisible $ \bigcup o_{t,i}[k]$ (operation informed in \cref{sec:comm_arch}) executed by the $i$-th agent at $t$, and $k$ denotes the operation index.
Following these definitions, the indivisible minimal events occurring within an LLMAS are transformed into operations $o$, which are bound to a specific time point, agent, and task(\textit{e.g.} perceive, think, act).
However, these low-level events can be irrelevant or redundant for high-level analysis targets. For a specific agent, there may exist hundreds of events at a single time point $t$, which imply factual (\textit{e.g.} duplicate segments generated by prompt construction) and semantic (\textit{e.g.} repeated biased interpretations of the same observation) duplications. 

To address these problems, we synthesize events on $T$ for each agent into their behaviors:
\begin{equation}
 B_{i,t_0\cdots t_1} =\bigcup_{t\in [t_0,t_1]}s_{t,i} 
 \label{eq:behavior}
\end{equation}
It refers to the set of operations performed by the $i$-th agent across the subsequence $[t_0,t_1]$ within the temporal series T.

}

\subsection{\note{Behavior} Summarization}
\label{sec:Event Summarize}
\begin{figure*}[!t]
    \centering
    \includegraphics[width=1\linewidth]{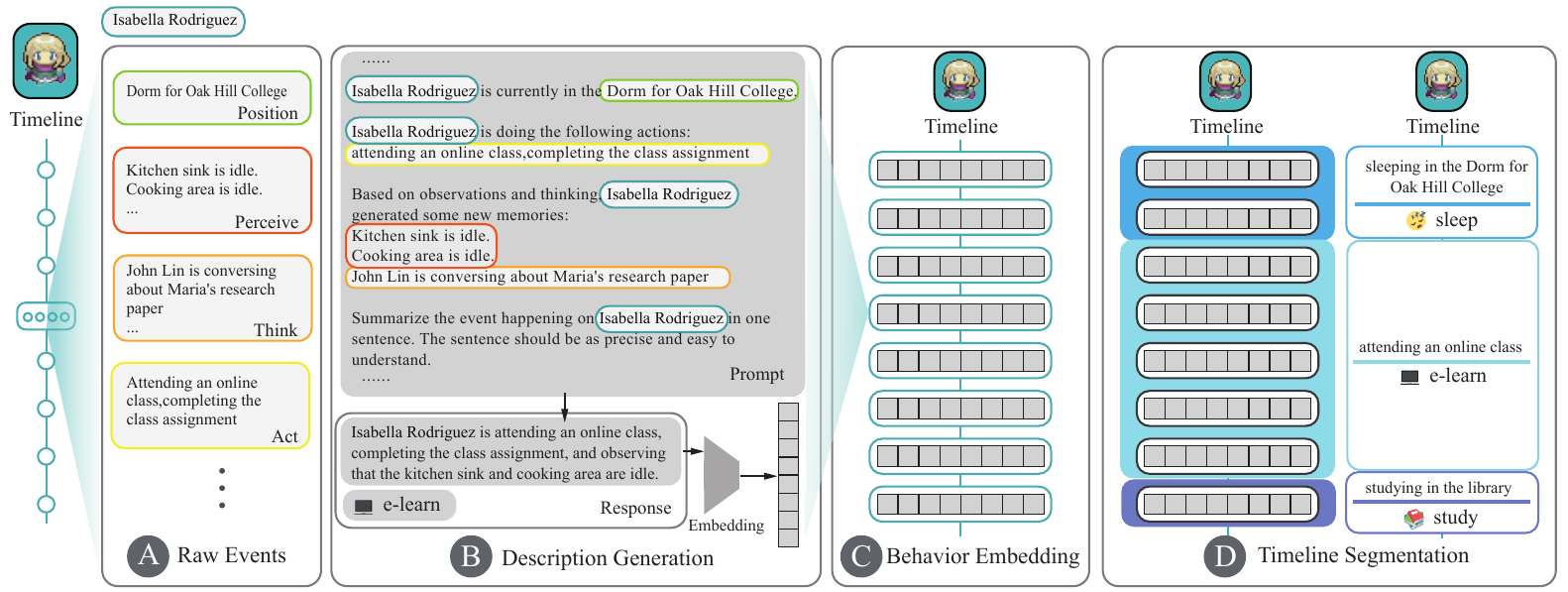}
    \caption{
    \note{The agent behavior is summarized in four stages: (A) Raw Events: acquire raw events from the logs to detail the occurrences involving the agent along the timeline, including the agent's location, actions, memory, and conversations. (B) Description Generation: organize the raw events and employ models such as LLMs to generate concise descriptions of the behaviors. (C) Behavior Embedding: translate the behavior descriptions into a sequence of textual embedding vectors. (D) Timeline Segmentation: involve the detection of change points within the sequence of behavior vectors, followed by the corresponding segmentation of the agent's timeline.}}
    \label{fig:summary}
\end{figure*}

In various LLMAS, operations manifest in different forms, such as text, images, and even physical behaviors in the factory environment.
\note{Meanwhile, new behaviors of those agents are continuously generated as $T$ is increasing. The multiplicity of manifestation and the extensive aggregation of behaviors can obscure the visualization system's interpretation}, thereby impeding the exploration of an agent's internal causality.
Therefore, we propose a behavior summarization method.
\note{
As shown in \cref{fig:summary}, we  (1) outline behaviors that encapsulate a singular time point into a succinct description (\cref{fig:summary}, \mybox{A} $\to$ \mybox{B}), (2) utilize text embedding to capture underlying semantics within the behavior(\cref{fig:summary}, \mybox{B} $\to$ \mybox{C}), (3) utilize a change point detection method to divide the sequence of behaviors and abstract each sub-sequence of behavior(\cref{fig:summary}, \mybox{C} $\to$ \mybox{D}). Ultimately, we can summarize a multitude of small behaviors into several noteworthy behaviors with segmented timelines.}

\note{
\textbf{Description Generation:}}
We incorporate an external text summarization model, which acts as a standalone LLM agent that operates independently of LLMAS. 
All annotated descriptions are concatenated to form a comprehensive model input (\textit{i.e.} prompts for LLM). Given this long text sequence as input, the summarization model generates a succinct behavior description, significantly reducing the information length while maintaining the original meaning \note{(shown in \cref{fig:summary}, \mybox{B}, from \textit{Prompt} to \textit{Response})}. 
Concurrently, we prompt that the summarization model yields a highly abstract description of the behavior, employing both textual and emoji symbols. Textual descriptions serve as the foundation for the forthcoming embedding model and emoji symbols are conceived to facilitate subsequent visualization.

\note{\textbf{Behavior Embedding:}
We further utilize all summarized behavior descriptions, embedding them to better grasp the latent semantics, including the inherent similarities and hierarchical relationships.}
To maximize the efficiency of the encoding schema, we adopt the text-embedding model\footnote{https://platform.openai.com/docs/guides/embeddings} pretrained on large-scale internet text data, renowned for its superior performance, cost-effectiveness, and simplicity of use. 
\note{The summarized behavior descriptions are then each encoded into a 1536-dimensional vector, constituting the sequence $E_{\text{agent}}$ for each agent. }
With these powerful embeddings, we can uncover the semantic similarity of a single behavior, thereby unlocking the potential to tackle a myriad of complex text sequence analyses.

\note{\textbf{Timeline Segmentation:}}
Considering the data characteristics of the embedding sequence $e$ and our design requirements, we employ the Window-based change point detection (WIN) algorithm \cite{chu1995time} with the cosine distance measure to segment the sequence. \note{This approach is suitable for real-time or streaming data contexts, as it allows for incremental updates in response to the arrival of new data and exhibits insensitivity to short-term and frequent fluctuations.}

Firstly, to compare two embedding vectors $e_x$ and $e_y$  ($e_x,e_y \in E_{\text{agent}}$) with dimension $d=1536$, we use the cosine similarity 
$k_{cosine}: \mathbb{R}^d \times \mathbb{R}^d \rightarrow \mathbb{R}$ (shown in \cref{equation:costcosine}) as the kernel function\cite{truong2020selective}
, where $\langle \cdot,\cdot \rangle$ and $\| \cdot \|$ are the Euclidean scalar product and norm respectively:
\begin{equation}
k(e_x,e_y):=\frac{\langle e_x|e_y \rangle}{\|e_x \| \|e_y \|}
\label{equation:costcosine}
\end{equation}

Then we recall the cost  $c(\cdot)$ deriving from $k(\cdot,\cdot)$  as \cref{costfunction}, where $e_{a..b}$ is the subsequence $\{e_{a+1},e_{a+2},\cdots,e_{b}\}\subseteq E$:
\begin{equation}
    c(e_{a..b})=\sum_{t=a+1}^b k(e_t,e_t)-\frac{1}{b-a}\sum_{s,t=a+1}^b k(e_s,e_t)
    \label{costfunction}
\end{equation}

WIN utilizes two sliding windows that traverse the data stream. By comparing the statistical properties of the signals within each window, a discrepancy measure is obtained based on the cost function $c$:
\begin{equation}
    d(e_{u..v},e_{v..w})=c(e_{u..w})-c(e_{u..v})-c(e_{v..w})
    \label{equation:discrpancy}
\end{equation}

The discrepancy $d$ means the cost gain of splitting the sub-sequence $e_{u..w}$ at the index $v$. If the boundary $v$ is a change index within the window $u..w$, the discrepancy $d$ will be significantly higher. After a sequential peak search of $d$, we have a series of time points $t_1^*<t_2^*<...<t_K^*$. Certain features of the embedding sequence change suddenly at these points. We utilize the abstraction of $t_i$, encompassing both textual and emoji symbol descriptions, to aggregate the behaviors of the agent from $t_i$ to $t_{i+1}$.

\begin{figure}[!h]
    \centering
    \includegraphics[width=\linewidth]{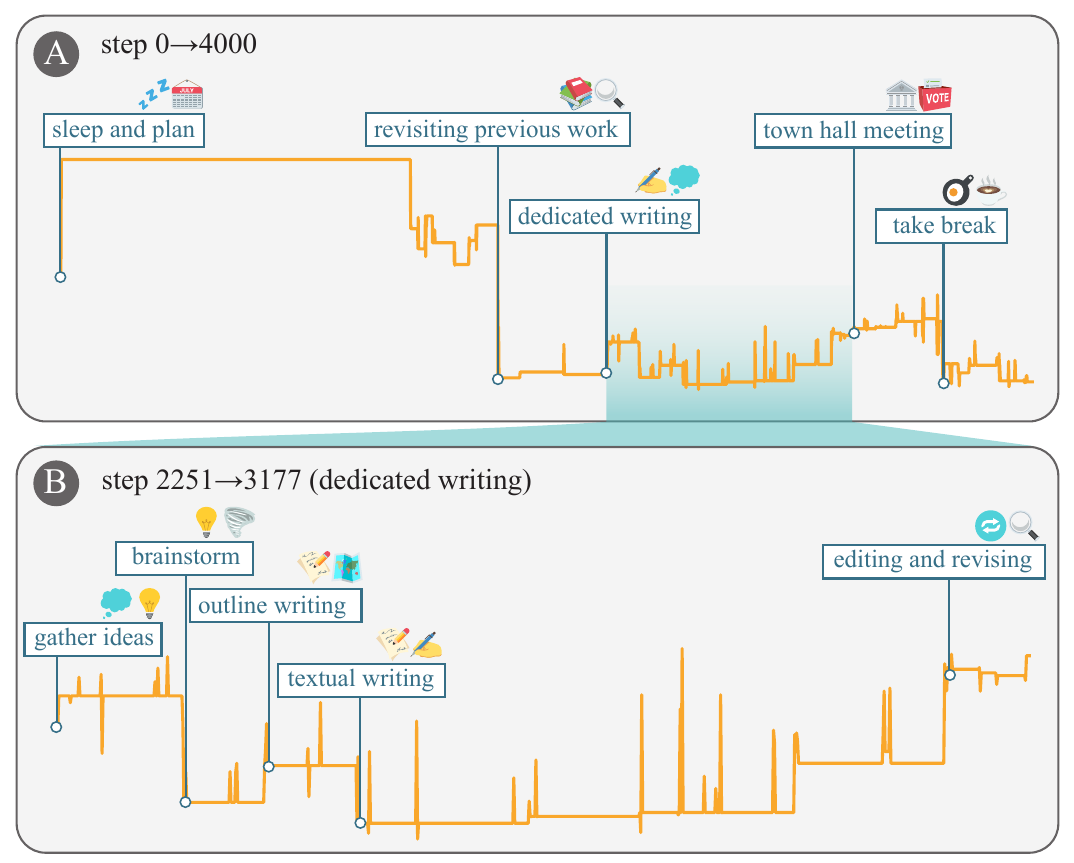}
    \caption{\note{The timeline segmentation results of an agent in Reverie. The x-axis denotes the timeline, while the y-axis corresponds to the value of the principle PCA component of agent behavior embedding at each time point.}}
    \label{fig:segmentation}
\end{figure}

\note{\cref{fig:segmentation} provides an illustrative example of the timeline segmentation process. Here we try to segment the timeline of a writer agent in the Reverie environment. The agent’s entire morning schedule is shown in ( \cref{fig:segmentation}, \mybox{A}), spanning from midnight to noon, encompassing 4000 time points (0$\to$4000) on the timeline. To facilitate an intuitive understanding of the segmentation result, we conducted principal component analysis (PCA) on the embedding of behavior at each time point, and used the y-axis to encode the values of the primary PCA components, resulting in the orange line plot presented in \cref{fig:segmentation}. 

As we can see in ( \cref{fig:segmentation}, \mybox{A}), by applying the segmentation algorithm (with N=5 as an example), this period is summarized into five main behaviors (``sleep and plan'', ``revisiting previous work'', etc.). Moreover, if we re-apply the timeline segmentation algorithm to the ``dedicated writing'' behavior, which spans time points 2251$\to$3177 (\cref{fig:segmentation}, \mybox{B}) on the timeline, we can further divide it into five sub-behaviors (``gather ideas'', ``brainstorm'', etc.). Note that all these sub-behaviors can be considered as ``dedicated writing'', while exhibiting more subtle distinctions among them.

Another observation to note is that in the line plot formed by PCA principal component values, there are some peaks. These peaks occur because the agent executes specific operations at this time point, such as generating new memories or perceiving new objects. However, these operations do not have a lasting impact on the agent’s ongoing behavior. Therefore, they are usually regarded as tiny behaviors contained in their parent behavior.}

\subsection{Cause Tracing}
\label{sec:CauseTrance}

Within a complex timeline, any agent event is influenced by both its internal memory and interactions with the external environment. By tracing the causal factors of these events, users can gain valuable insights into agent behaviors (\textbf{R3}) and LLM invocation for decision-making (\textbf{R4}), thereby improving the credibility and interpretability of LLMAS.

\note{
Existing works\cite{park2023generative} primarily rely on log debugging to explicitly reveal the origins of agents' operations. However, these methods place an additional cognitive burden on users due to the need for manual tracing and often fail to capture implicit causal relationships. For instance, current thinking can be influenced by observations over a long time steps.}
To efficiently trace the behavior causes, we propose a two-fold provenance tracing method to mine the causal relationships between underlying events within the behaviors. 

\note{
\textbf{Explicit Causes: }
It refers to the distinct and observable causal relationships that can be directly discerned from raw event logs, explicitly delineating the direct influence relationships between operations. For example, in open-source agent creation frameworks like Langchain\cite{Langchain} and AgentVerse\cite{chen2023agentverse}, mechanisms have been implemented to index attributes of agent memory, facilitating direct backtracking to the relevant source operations upon the invocation of an agent’s memory. When such explicit causal chains are completed in LLMAS, users can thus obtain these records through raw event logs and transmit them to \sysname{}. \sysname{} utilizes these logs as input to facilitate the analysis of downstream tasks for users.
}

\note{
\textbf{Implicit Causes: }
Throughout the evolution of LLMAS, the agents’ invocations of historical operations are not always documented, but rather are expressed through complex intermediate variables or latent patterns within the program. To capture these implicit causal relationships, we conduct relevance detection based on the text similarities (as in \cref{equation:costcosine}) between the textual log of these operations themselves, thereby revealing the latent connections between events. To strike a balance between uncovering potential causal relationships and preventing information overload for users, we define a similarity threshold $\delta$. For a certain operator $o_{res}$ at time point $j$, if the similarity between it and another operator $o_{src}$ at time point $i$ (\textit{s.t.} $i\leq j$) exceeds $\delta$, we consider $o_{src}$ as one of the potential causes of $o_{res}$.
}

\note{
After the extraction of both explicit and implicit causes among operations is completed, we have ascertained every possible pair $<o_{src},o_{res}>$. The connections between operations can be elevated to the connection between the corresponding behaviors in a bottom-up fashion, in accordance with the definition of behavior outlined in \cref{sec:Behavior Definition}.
}
\section{User Interface}
\label{sec:interface}
The user interface is composed of three views. The \textit{Outline View} (\cref{fig:teaser}, \mybox{A}) visualizes how the agents' activity, interaction, and environment change over time, allowing users to analyze the evolution process of the LLMAS. Once the user becomes interested in certain behaviors of any agent, they can check its details and trace its cause from the \textit{Agent View} (\cref{fig:teaser}, \mybox{B}). During the exploration process, the visualization of LLMAS will synchronously switch to the corresponding agent and time point to support intuitive perception and verification in \textit{Monitor View} (\cref{fig:teaser}, \mybox{C}).

\subsection{Outline View}

\textit{Outline View} serves as a springboard for exploration, providing a suitable generality of information (\textbf{R1}) to assist users in efficiently discovering noteworthy patterns or behaviors of interest during the evolution of the LLMAS. 

\textbf{Agent Timeline Summarization:} Every agent has its individual behaviors (\textit{e.g.} what it is perceiving, thinking, and acting) at each time point. When users double-click on the view, all selected agent curves will be automatically summarized into N (we set N = 10 during experiments) segments using the behavior summarization algorithm proposed in Section \ref{sec:Event Summarize}. Users can click the start of a segment to check details about what is happening during this period of timeline. If users desire a more granular behavior representation (\textbf{R1}), they can zoom in to a specific region by scrolling the mouse wheel. The system will then re-summarize the timeline based on the currently visible area (\cref{fig:teaser}, \mybox{$A_1$}).

\textbf{Agent Interaction Analysis:} Each agent in the \textit{Outline View} is represented as a uniquely colored curve, whose x-axis encodes the system time point and y-axis encodes the location of the agent, depicting the transition of the location of each agent (\textbf{R2}). When several agents are in the same time and location, they can have interactions (\textit{e.g.} conversations, collaborations, or conflicts) with each other. Since these interactions usually play a crucial role in affecting the LLMAS’s evolution, we highlight them by filling the area among the corresponding segment of agent curves. Users can click an interaction area of interest to check the integration details (\cref{fig:teaser}, \mybox{$A_2$}). Drawing inspiration from previous work of storytelling\cite{10.1145/1879211.1879219,6327274}, we enforce agent curves to get closer if there is an interaction among them.

\textbf{Agent Memory Search:} Sometimes users want to conduct exploration about when and how the agents start to have thoughts about a specific topic (\textbf{R2}). Therefore, we provide a search box in the top right corner of the view, allowing users to add keywords related to the topic they want to explore. Whenever a keyword is added, the points on the agent curves corresponding to time points associated with relevant memory will be highlighted (\cref{fig:teaser}, \mybox{$A_3$}).

\subsection{Agent View}
When users notice a specific phenomenon or behavior from the \textit{Outline View} and wish to further explore it, they can click on the corresponding time point on an agent curve to access more details (\textbf{R1}) in the \textit{Agent View}.

\textbf{Agent Characteristic:}
A complex LLMAS typically contains agents with different characteristics. For example, agents might be assigned different roles and goals, which are usually realized through prompt engineering or LLM fine-tuning. Since these details are important for users to understand and infer an agent's behavior, we display them on the left panel of the \textit{Agent View} (\cref{fig:teaser}, \mybox{$B_1$}).

\textbf{Time Point Revealing:}
On the right panel of \textit{Agent View}, we provide users with a timeline (\cref{fig:teaser}, \mybox{$B_2$}) to help investigate the behavior of the selected agent during this period of time, which is a detailed counterpart of the agent curve in \textit{Outline View}. Users can click a time point icon to reveal descriptions (summarized using the method shown in \cref{fig:summary}, \mybox{A}-\mybox{B}) and the task-level events performed by this agent at this time point (\textbf{R1}). They can click a task icon to further reveal the operators involved in performing this task (\textbf{R1}). As discussed in Section \ref{sec:comm_arch}, the operators can be classified into Environmental Operations, Memory Operations, and Decision Operations based on the target. Therefore, we use different icons to represent operators of different type: If the user clicks \raisebox{-0.2\height}{\includegraphics[height=1em]{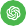}}, a description panel will pop up to show the invocation context of LLM to make the decision (\cref{fig:teaser}, \mybox{$B_3$}) (\textbf{R4}); If the user clicks  \raisebox{-0.2\height}{\includegraphics[height=1em]{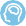}}, a description panel will pop up to show the texts stored into the memory at this operation; If the user clicks  \raisebox{-0.2\height}{\includegraphics[height=1em]{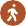}}, a description panel will pop up to show what the agent is perceiving from or act on the environment. 

\textbf{Cause Tracing:}
In addition to obtaining detailed behavioral information about agents, users also need to locate and analyze the reasons behind these agent behaviors. Whenever the user clicks an operator icon in the \textit{Agent View}, the system will utilize the cause trace method described in Section \ref{sec:CauseTrance} to find previous operators that potentially have an intrinsic relationship with the current operation and highlight their corresponding time point on the \textit{Agent View} (\textbf{R3}). We use edges with orange color to connect the selected operator and their predecessors.
Since the agent behaviors could be affected by previous operations a long time ago, we provide users with a mini-map to visualize the point of the current operation and its related predecessors across the whole timeline (\cref{fig:teaser}, \mybox{$B_4$}) (\textbf{R1}). Based on this mini-map, users can switch back and forth between the cause and result across the timeline more easily.

\subsection{Monitor View}
LLMAS typically provides a graphical representation of the dynamic simulation. It could be re-playable for 2D video or 3D, contingent upon the LLMAS evolution logs provided by the user for \sysname{}. This visual representation transforms abstract simulation data into perceptually friendly visual elements, which helps users understand LLMAS and verify their analysis more intuitively. However, manually switching between different locations and time points can be tedious and interrupt the user's analysis flow. \note{Therefore, we provide the \textit{Monitor View} to support fluent adjustment of the panoramic visualization of LLMAS (\cref{fig:teaser}, \mybox{C}) based on users' current focus and demand for context.}

\note{\textbf{Focus Switching:} Whenever the user clicks a time point on agent curve from the \textit{Outline View} or a time point from the \textit{Agent View}, the \textit{Monitor View} will automatically switch to the location of that agent at that time point, providing a corresponding concrete visualization to complement the other two views (\textbf{R1}).

\textbf{Context Revealing:} The \textit{Monitor View} also supports spatial and temporal context revealing to help users better comprehend the current focus point. As for the spatial context, the user can scroll the mouse wheel to adjust the level of scope, ranging from a macroscopic view of the entire LLMAS to a microscopic focus on a single agent. As for the temporal context, whenever the user changes the focus point from time point A to time point B, they can right-click the mouse to replay a fast-forward recording of that period of time in the \textit{Monitor View}. }

\section{Usage Scenarios}

\subsection{Scenario A: Information Diffusion}
\begin{figure}[h!]
    \centering
    \includegraphics[width=1\linewidth]{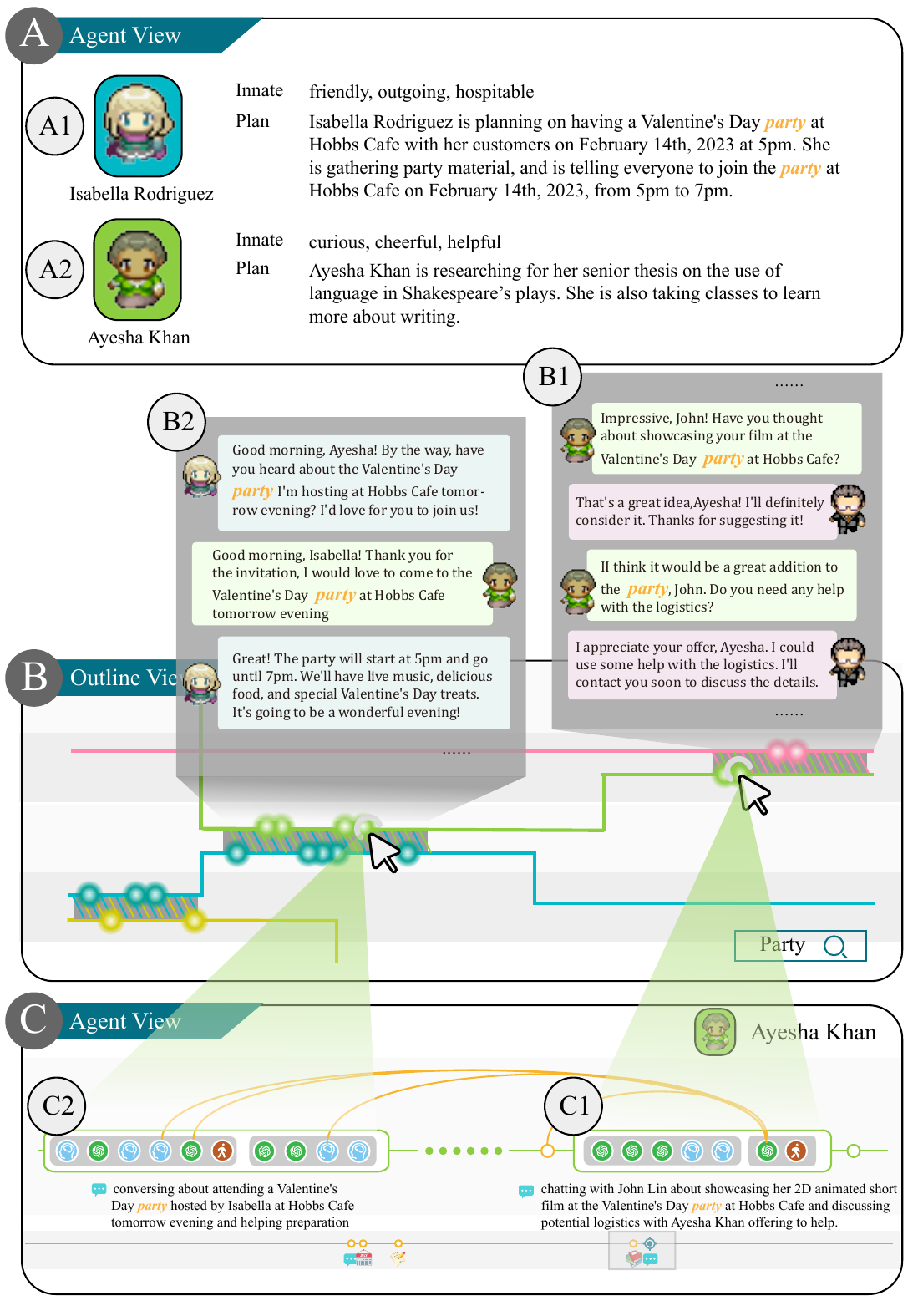}
    \caption{The first usage scenario showcases the support \sysname{} provides to the user in exploring social patterns like \textbf{Information Diffusion}.  (A) The user gleans the characteristics of each agent via the \textit{Agent View}. (B) Proceeding to the \textit{Outline View}, the user searches for the keyword ``\textit{party}'', discovering several related memory points generated in several conversations. (C) Utilizing the \textit{Agent View}, the user delves into the origins of these conversational patterns.}
    \label{fig:case2}
\end{figure}

This case demonstrates how our system helps users understand the patterns of agent behaviors in LLMAS.
In the initialization phase, the user adds the information ``\textit{Organize Valentine’s Day party at Hobbs Coffee on the evening of February 14th}'' to the characteristic (\cref{fig:case2}, \mybox{$A_1$}) of the agent Isabella Rodriguez (IR) and wishes to observe the evolution of the system on February 13th. 

To focus on the theme of the party, the user searches for the occurrence of the keyword ``\textit{party}'' (\cref{fig:case2}, \mybox{B}) in the agent’s memory and follows IR’s timeline for observation. The user discovers that the message primarily spreads during IR’s conversations with others.
Furthermore, the user finds the ``\textit{party}'' memory highlight surfacing in the conversation between Ayesha Khan (AK) and John Smith (JS). Upon examining their dialogue (\cref{fig:case2}, \mybox{$B_1$}), it is revealed that the message is from AK to JS while there is no prior knowledge of the ``\textit{party}'' message in AK's settings (\cref{fig:case2}, \mybox{$A_2$}). 
In order to delve into the underlying cause, the user selects the time point when AK initiates the conversation with JS, employing the \textit{Agent View} to obtain detailed insights (\cref{fig:case2}, \mybox{C}).
The user expands the time point (\cref{fig:case2}, \mybox{$C_1$}) and traces the cause of one of the decision operations (\cref{fig:case2}, \mybox{$C_2$}). It is highly probable that AK's decision to discuss ``\textit{party}'' with JS has its historical roots in a conversation between IR and AK that took place some time ago.
Finally, the user reverts to the \textit{Outline View}, confirming that a conversation concerning the ``\textit{party}'' has indeed occurred between IR and AK (\cref{fig:case2}, \mybox{$B_2$}), during which IR extends an invitation to AK to participate in the party preparation.

With the assistance of \textit{AgentLens}, the user successfully pinpoints an instance of information diffusion from a primary disseminator IR to a secondary one AK, then gradually diffusing towards other agents. From the \textit{Agent View}, users discover that with the increase in both secondary propagators and the number of conversations related to ``\textit{party}'', the speed of ``\textit{party}'' diffusion throughout the small town significantly accelerates.

\subsection{Scenario B: Unexpected Social Patterns}

\begin{figure}[h!]
    \centering
    \includegraphics[width=1\linewidth]{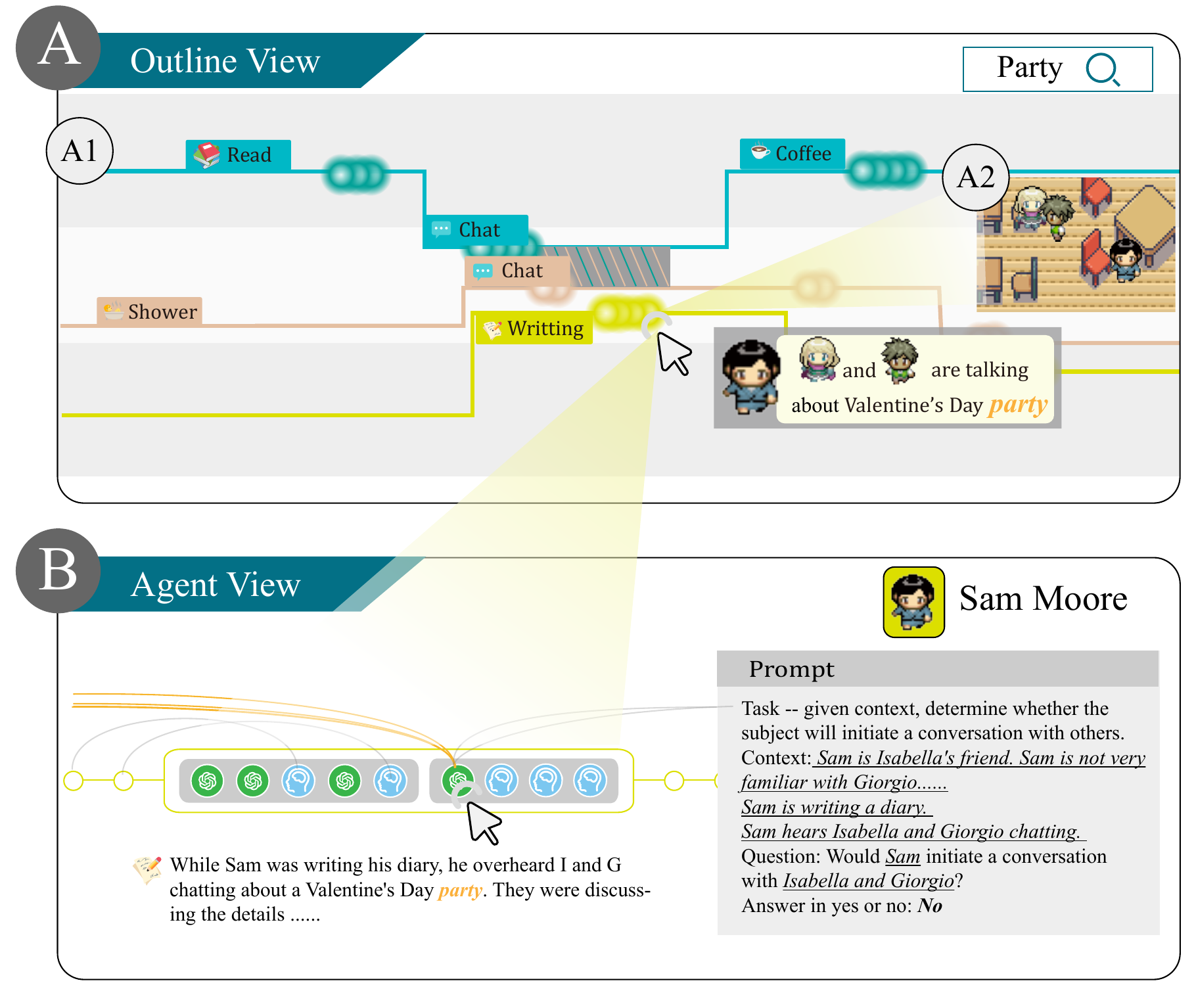}
    \caption{The second usage scenario presents how \sysname{} aids users in explaining an unexpected agent behavior. (A) The user identifies some unexpected agent behaviors in \textit{Outline View}, like an agent participating in information dissemination without engaging in a related conversation. Upon validation through \textit{Monitor View}, the user determines that this pattern corresponds to the eavesdropping behavior of the agent. (B) The user uses \textit{Agent View} to investigate the reasons behind the agent's reluctance to participate in the discussion. Finally, the user discovers that a certain decision operation at the time point results in the behavior.}
    \label{fig:case1}
\end{figure}

In this scenario, the user uncovers an unexpected pattern of information diffusion: \textbf{eavesdropping.}

During the observation of the ``\textit{party}'' propagation process (\cref{fig:case1}, \mybox{$A_1$}), the user discovers that Sam Moore (SM) forms relevant memories without engaging in any direct conversation. 
According to the event summary, SM is in the process of writing his novel when this memory is formed. The user hovers on this memory point about the ``\textit{party}'', learning that the memory formed by SM at this time is ``\textit{IR and Giorgio Moore (GM) are talking about Valentine's Day party}''. From the visual representation, the user observes that IR, SM, and GM are in the same room at this moment, a fact that is corroborated by the \textit{Monitor View} (\cref{fig:case1}, \mybox{$A_2$}). The user infers that SM comes to know about the ``\textit{party}'' by eavesdropping on others' conversations.

The user seeks to investigate why SM does not join the conversation. The user expands the corresponding time point(\cref{fig:case1}, \mybox{B}) in the \textit{Agent View} and identifies the \textit{Decision Operation} that determines SM's choice not to participate in the discussion.
The prompt dispatches to the LLM incorporated agent settings pertaining to SM, like ``\textit{SM is IR's friend}'' and ``\textit{Sam is not very familiar with GM}'', in addition to the immediate observations made by SM, such as ``\textit{IR and GM are presently engaged in a conversation}'' among other pieces of prompt input. It is the response returned by the LLM, based on the prompt, making the decision for SM's subsequent action that he determines not to join the conversation.

\section{User Evaluation}

We conducted a user study to evaluate the performance of \sysname{} in enhancing LLMAS analysis. The study was specially designed to assess the comprehensive efficiency, effectiveness, and usability of the system. We also examine the analytical support provided by our system compared to a baseline system, which replicates the visual approach in existing LLMAS works.

\subsection{Participants}

\note{To prevent participants from having prior knowledge of the system before evaluation, we recruited 14 new participants (denoted as P1-P14) from a local university who had not been involved in the design requirements phase of this study, thereby enhancing the assessment validity and the results generalizability. }These participants have diverse academic backgrounds, with most being undergraduate and graduate students from fields such as computer science, software engineering, and sociology. Some of them are developers with a high level of expertise in LLMAS, while others only have had direct interaction with LLMAS.

\subsection{Baseline Systems}
A baseline system\footnote{https://reverie.herokuapp.com/arXiv\_Demo/\#} has been set up for direct comparison with our proposed system. Both the baseline system and our system utilize the log data generated by Reverie\cite{park2023generative}, which records the interactions and memory logs of agents within the system during the simulation process.

The baseline provides a view for replaying past events with plain text descriptions of agent settings and behaviors, which simulates a typical LLMAS panoramic visualization. Firstly, it features a monitoring interface that uses a flat map as the background. This allows users to replay and observe the agent positions and behavior descriptions at different time points through a timeline. Secondly, the system offers a textual representation of the current events for each agent, including the agent's location, the action in progress, and the ongoing dialogue (if any). Finally, the system also provides a pure textual display of all events in each agent's evolutionary process, 
encompassing the agent's personality, complete memory records, and event sequences.
These features enable users to understand the agent behaviors and status and delve into their evolutionary process.

\subsection{Procedure and Tasks}

\textbf{Introduction (10 min): }
Initially, we provided a concise overview of the research, including the motivation and methodology. We then collected basic personal information from them, including their gender, age, and occupation. In addition, we obtained authorization to record their behaviors during the subsequent task analysis. Finally, we describe the characteristics of the individual views in both baseline and \sysname{} in detail and demonstrate their practical use in a specific scenario.

\textbf{Task-based analysis (40 min): }
\note{In this stage, participants were required to undertake 2 groups of analytical tasks (refer to \cref{fig:evaluation_t1,fig:evaluation_t2}), designed to evaluate the system's overall effectiveness and usability. }Participants were required to fulfill tasks for each system, with the duration and accuracy of task completion being recorded. 
To obviate the potential for participants to replicate responses through memorization \cite{feng2023xnli}, the sequence in which the two systems were presented was randomized. Each task was uniquely tailored for both systems while ensuring an equivalent level of challenge.

\textbf{Semi-structured interview (30 min): }
To enhance the evaluation of the method and interface efficacy, we utilized the five-point Likert scale in an 8-item questionnaire. \note{Additionally, we employed the System Usability Scale (SUS)\cite{brooke@sus} to evaluate the usability of \sysname{}. }Participants were asked to rate each question from 1 (strongly disagree) to 5 (strongly agree)  to gauge their agreement levels.
During the questionnaire process, we encouraged participants to speak freely to uncover the reasoning behind their ratings.

\subsection{\note{Task Completion Analysis}}

For the task-based analysis, we conducted a quantitative comparison between \sysname and the baseline, focusing on accuracy and task completion time. 
\note{We developed two distinct groups of evaluation tasks to assess the efficacy of 2 systems for the analysis of agent behaviors (\cref{fig:evaluation_t1}) and the identification of emergent phenomena arising from such behaviors (\cref{fig:evaluation_t2}).}

\subsubsection{\note{Individual Behavior Analysis}}

\begin{figure}[h!]
    \centering
    \includegraphics[width=1\linewidth]{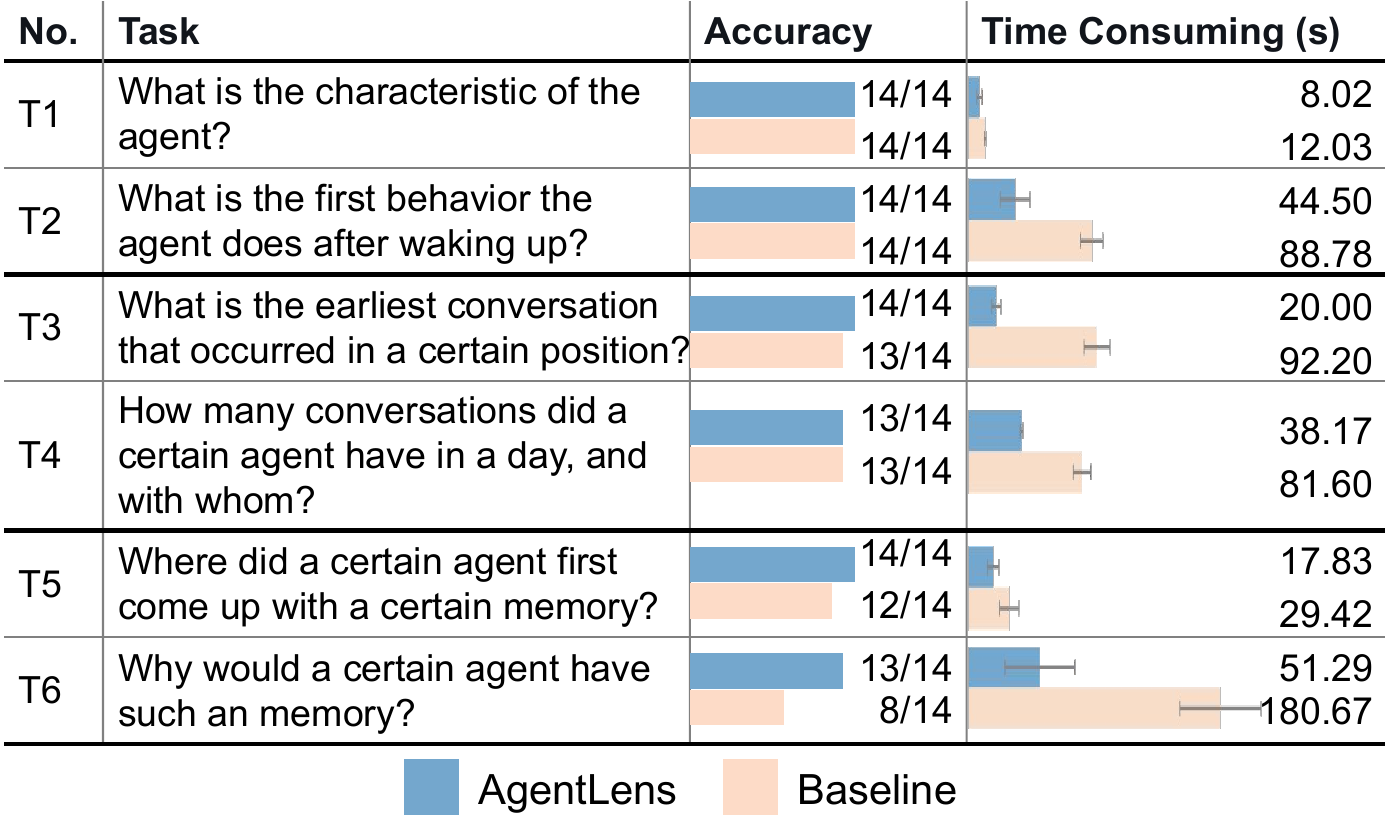}
    \caption{Statistical result of the accuracy and time consumption for participants completing individual behavior analysis tasks using both \sysname{} and the baseline system.}
    \label{fig:evaluation_t1}
\end{figure}

\note{T1 - T6 in \cref{fig:evaluation_t1} are designed with elicit concise answers, requiring participants to rapidly comprehend the fundamental characteristics and behaviors of agents.}
Based on the analytical target, we categorize this set of tasks into 3 classifications.
Participants exhibit varying levels of accuracy and time expenditure across tasks, however, there was a notable improvement in task accuracy ($p=1.2e-3$) and reduction in time consumption ($p=1.2e-3$) with \sysname{}.

\note{\textbf{Single-agent analysis (T1 - T2): }
This set of tasks focuses on the system's enhancement of simple information analysis about individual agents.}
Without compromising task accuracy, \sysname{} decreased time consumption by 33\% for T1 ($\mu_{AgentLens}=8.02,\mu_{baseline}=12.03$) and by 50\% for T2 ($\mu_{AgentLens}=44.50,\mu_{baseline}=88.78$) compared to the baseline system.
The visual representation of agent characteristics in the \textit{Agent View} eliminates the need for search operations in T1. Furthermore, the event summarization method helps participants quickly identify agent behaviors, eliminating the need to sift through complex log records to complete T2.

\note{\textbf{Multi-agent analysis (T3 - T4): }
This set of tasks demonstrates the system's effect in assisting participants with the analysis of interactions between agents.}
It is noteworthy that one participant failed in both two tasks using the baseline system due to his incorrect agent selection.
\sysname{} reduced time consumption by 78.3\% for T3 ($\mu_{AgentLens}=20.00,\mu_{baseline}=92.20$) and 53.2\% for T4 ($\mu_{AgentLens}=38.17,\mu_{baseline}=81.60$). The visual encoding in \sysname{}, particularly in the \textit{Outline View}, allowed participants to quickly derive answers by observing agent interactions including dialogues and cohabitation instances.

\note{\textbf{Behavior Cause analysis (T5 - T6): }
In this set of tasks,  \sysname{} demonstrated marked improvements over the baseline in facilitating the exploration of the cause of agent behaviors.}
While a part of the participants quickly obtained answers using the baseline in T5, \sysname{} still provided a 39.4\% improvement with the topic search feature ($\mu_{AgentLens}=17.83,\mu_{baseline}=29.42$).
T6 presented a significant challenge for the baseline, with over 42\% of participants notably failing to complete the task.
P9 commented, ``\textit{In the ton of plain text logs, I can't find any connection between the events at all.}''
However, with the cause trace feature in \textit{Agent View}, \sysname{} demonstrated a substantial 71.6\% improvement in it ($\mu_{AgentLens}=51.29,\mu_{baseline}=180.67$).

\begin{figure}[h!]
    \centering
    \includegraphics[width=1\linewidth]{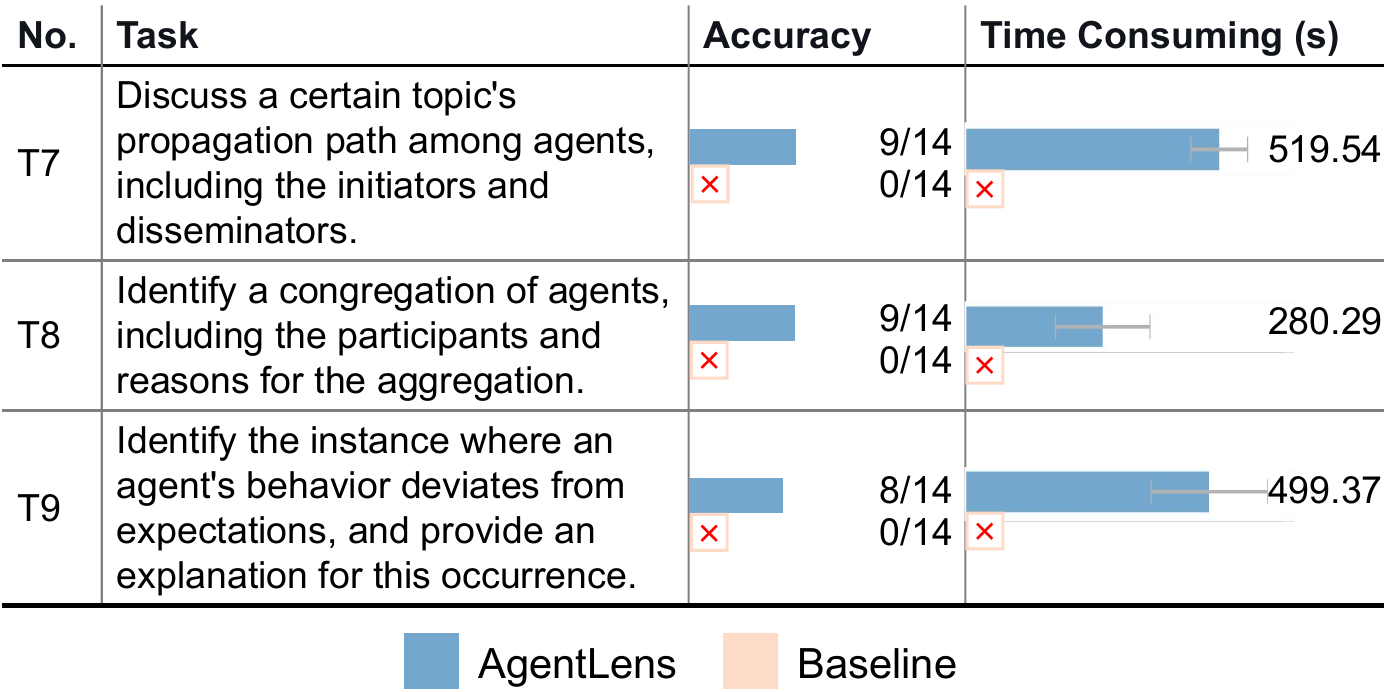}
    \caption{\note{Statistical result of the accuracy and time consumption for emergent phenomena identification tasks using both \sysname{} and the baseline system.}}
    \label{fig:evaluation_t2}
\end{figure}

\note{
\subsubsection{Emergent Phenomena Identification}

T7-T9 in \cref{fig:evaluation_t2} are designed to correspond to three categories of emergent phenomena arising from agent autonomy, which is not explicitly pre-programmed in LLMAS. These tasks are more complex for the participants, requiring back-and-forth exploration and analysis through multiple steps. We invited evaluators to assess the accuracy of the participant's responses.
Concurrently, we observe that \sysname{} demonstrates capabilities in complex analytical tasks that the traditional baseline failed to achieve, particularly in the exploration of emergent behaviors arising from agent autonomy.
}

\note{\textbf{Topic propagation (T7): }
Participants are tasked with identifying the propagation path of a specific topic, such as ``a Valentine's Day party will be held'' or ``someone is preparing the selection for mayor''.
Nearly all participants consider the task to be impossible while utilizing the baseline, as ``this task is akin to searching for a needle in a haystack'' (P11).
When utilizing \sysname{}, the majority of participants swiftly opted for the \textit{Agent Memory Search} within the \textit{Outline View} to conduct searches on the propagated topics.
Leveraging the representation of \textit{Agent Interaction Analysis} within the view, participants could easily explore the propagation paths.
Although the propagation path participants were asked to identify has multiple branches and complex scenarios, 9 participants completed the task using \sysname{}.
}

\note{\textbf{Agent congregation (T8): }
Participants are required to identify a congregation phenomenon, defined as more than three agents engaging in the same behavior at the same location, and participants should explain the reason behind it.
While using the baseline, participants were compelled to conduct extended observations and iterative replays of the recorded video. Despite locating the participants of the aggregation, they remained unable to ascertain the underlying causes of the phenomena.
Through the interactivity among the three views of \sysname{}, particularly the design of \textit{Monitor View} and \textit{Outline View}, participants were able to rapidly detect aggregation phenomena. Coupled with the method of behavior summarization, 9 participants successfully provided explanations for the aggregations.}

\note{\textbf{Unexpected behavior (T9): }
Participants were tasked with identifying and rationalizing unexpected agent behaviors across two systems. When using the baseline system, they noted that agent behaviors appeared uniformly logical and coherent. Additionally, the requisite alternation between observing multiple agents hindered their analytical process, thereby increasing the difficulty of detecting unexpected phenomena.
With the assistance of \sysname{}, this task became more manageable. P5 identified through \textit{Outline View} that "agent RP did not leave his room throughout the entire day." He traced the cause using \textit{Agent View} and discovered that the agent had received a plan that did not require leaving the house from LLM during the planning phase for that day. Another participant P8 noticed in \textit{Agent View} that agent TT was able to observe the activities of agent IR in the adjacent room, and this observation influenced TT's subsequent decisions. The user suggested that this phenomenon should be addressed in the LLMAS, as in human society, individuals do not possess the ability to see through walls.}

\subsection{Semi-structured Interview Analysis}
\begin{figure}[h!]
    \centering
    \includegraphics[width=1\linewidth]{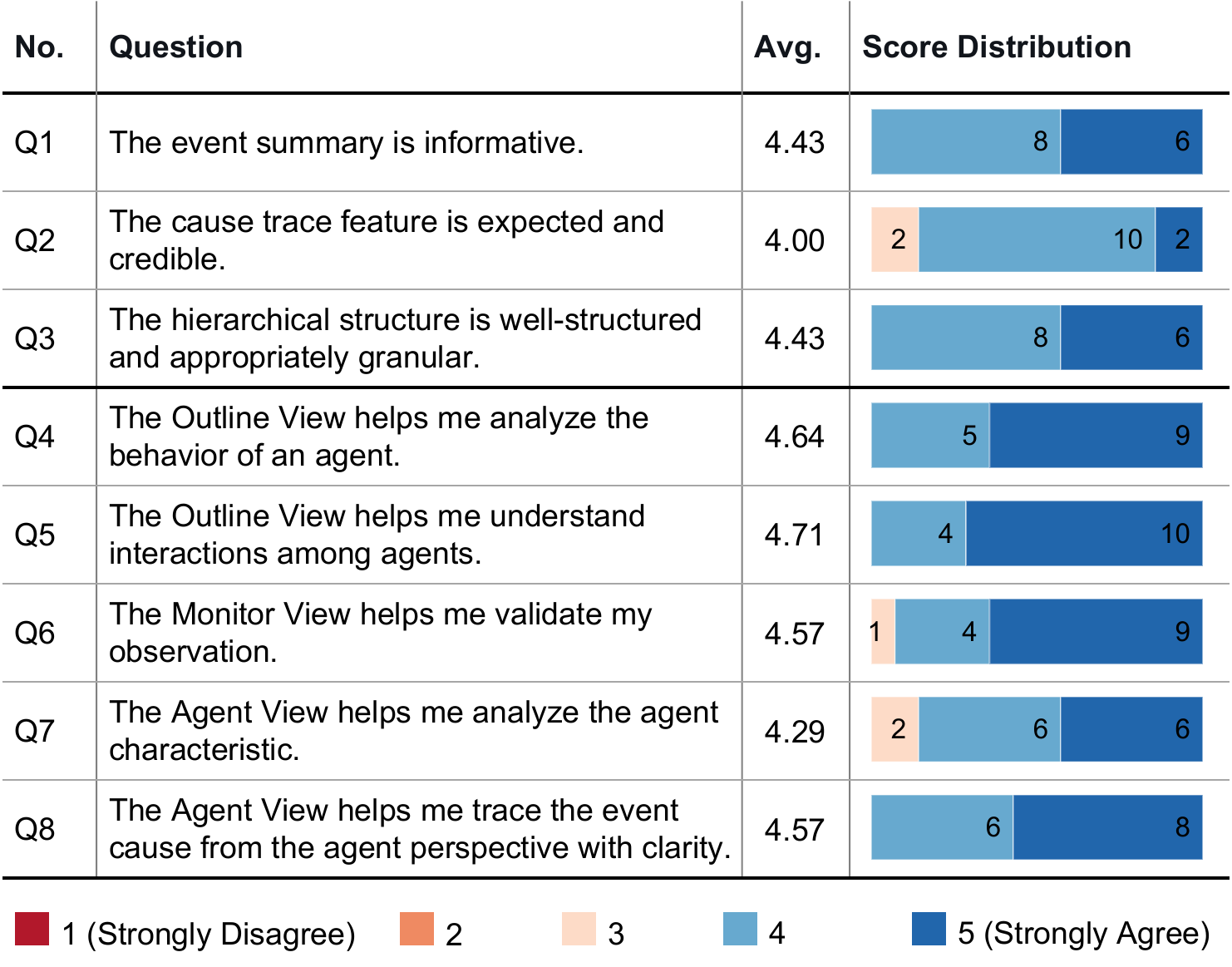}
    \caption{The questionnaire with results showing the efficacy of our method and interfaces.}
    \label{fig:evaluation2}
\end{figure}

\note{We posed 8 interview questions in \cref{fig:evaluation2}) and a SUS questionnaire(\cref{fig:sus}) to participants. 
Evaluating the results of the questionnaire with feedback obtained during the interview, we reported the performance of \sysname{} including its effectiveness and usability, offering insights into its practical application.
}

\subsubsection{Pipeline Effectiveness}
All participants agreed that the event summary is \textbf{informative} (Q1) and helpful. P10 commented, \textit{``The summaries are quite accurate. I can quickly locate the events and understand the evolution of an agent throughout the day with the help of the story-like subheadings.''} P1 felt impressed with the way of summarizing the agent's status, \textit{``like having an agent helping me monitor this LLMAS.''} 

Most participants agreed that the results of the \textbf{cause trace} met their expectations (Q2). They are willing to utilize the traced events to help analyze their interested events. For instance, P3 intended to incorporate the agent characteristics into the cause trace process. P5 pointed out that the cause trace served to \textit{``unveil the black box of agent behavior.''}. 

\textbf{The hierarchical structure} received unanimous endorsement from all participants (Q3). They all admitted that the hierarchical structure elucidated the level at which they could retrieve information. \note{Especially in the analysis of complex phenomena, the behavior hierarchical structure can \textit{``effectively reduce information density''}(P6) and \textit{``help me quickly focus on key phenomena''}(P10).} Nonetheless, P12, who was relatively inexperienced with the LLMAS, expressed a need for more \textit{``user-oriented guidance''}.

\subsubsection{Visual Effectiveness}
The \textit{Outline View} was appreciated by the participants for \textbf{agents behavior analysis} (Q4). It helps participants circumvent the risk of \textit{``getting lost in the complex and chaotic agent lines''} (P1) by summarizing and visualizing the agent's status. The interactive design, such as the click-to-highlight and view-details features, is \textit{``remarkably user-friendly and intuitive''} (P11). In addition, the encoding of \textbf{interactions among agents} also received positive feedback from users (Q5). The gray box, which intertwines two lines to represent agent dialogues, \textit{``stands out right away''} (P2). Some participants (P5, P7) indicated that they were accustomed to first spotting interesting agent dialogues in the relatively compact view, then zooming in to delve into more details. 
\note{P7, who completed the task of identifying congregation phenomenon(T8) expeditiously, attributes the success to \textit{"the visualization is trying to aggregate the curves of agents who are interacting with each other."}}
P9 commented, \textit{``If I can dynamically adjust the positions of the agents in the view, the layout can better match my expectation.''}

The \textit{Monitor View} was found to be useful for \textbf{validating the observation} (Q6). Several participants indicated that after observing the \textit{Monitor View}, they gained more confidence in the results of their analysis. P10 mentioned, \textit{``The monitor screen adjusts as I shift my focus in different views, kind of like video software, but it offers much more details than regular video playback.''}
\note{P10 commended the interaction of this view in relation to the other two especially in complex tasks, \textit{``This interactive responsiveness is beneficial during my iterative analysis process.''}}
P5 suggested that the \textit{Monitor View} could be more beneficial if it could \textit{``display the location information of other unfocused agents''}.

The \textit{Agent View} provides strong support for participants to analyze \textbf{individual agent characteristics} (Q7) and the \textbf{causal relationships} between agent behaviors (Q8). When observing agents of interest, they can \textit{``quickly understand the agent's personality and style of action''} (P1). P6 said, \textit{``The retrospective analysis is intuitive, but the individual timeline is too long. It would be better if I could explore the causes without having to drag the view around.''} P4 praised the minimap in the \textit{Agent View}, \textit{``When I was trying to understand agent behaviors, I love using the minimap's navigation. It helped me find the causal links fast with those cool summary emojis.''} P13 commented, \textit{``Developers should think about adding the agent view to their projects. Without it, agent behaviors might not seem convincing.''}

\subsubsection{Usability}
\begin{figure*}[h]
    \centering
    \includegraphics[width=1\linewidth]{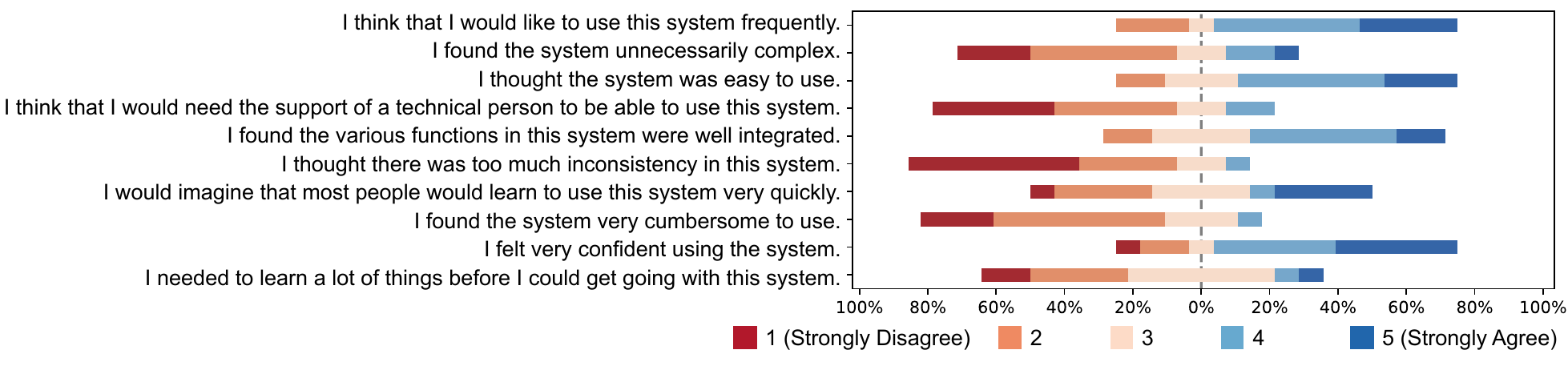}
    \caption{\note{The SUS questionnaire with results showing the usability of \sysname{}.}}
    \label{fig:sus}
\end{figure*}

\note{We employed the SUS questionnaire to assess the system usability, thereby reporting users' cognitive load with \sysname{}. 
Several developers among the participants conveyed not only their intent to use \sysname{} in the future but also to consider its integration within their LLMAS development, which has significantly encouraged us.
}

\note{
Overall, participants provided positive comments on the usability. P9 lauded the workflow of \sysname{}, \textit{``I thoroughly enjoyed the freedom of exploration the system facilitated.''}. 
P13 noted, \textit{``The interaction is very fluid''}, but revealed a longing for automated assistance during complex analytical tasks: \textit{``It would be perfect if the system could understand the type of task I want to analyze from just a few of my clicks.''}
Moreover, participants expressed their confidence and enjoyment when using \sysname{}. However, several participants indicated that the system necessitates a measure of preliminary technical knowledge, despite acknowledgment from P2 that \textit{``this is principally due to the intrinsic complexity of LLMAS itself.''}
}

\note{
Ultimately, we achieved an average score of 67.5 on the SUS questionnaire(refer to \cref{fig:sus}), which we find exhilarating. However, it also serves as a reminder of the necessity for future optimization. 
}

\section{Discussion}

In this section, we commence by encapsulating the lessons collected from the user feedback, 
including providing comparisons within an agent and enabling modifications for system configurations. 
Subsequently, we deliberate on the generalizability, as well as the limitations and future work.

\subsection{Lessons Learned}

\textbf{Providing comparison within an agent.}
During the evaluation process, we recorded some specific interaction patterns among the users, although they did not actively mention them in the interview.
Some users frequently analyzed the behaviors of a single agent across various temporal intervals. For instance, they compared the behaviors of an agent at 8 a.m. on February 13 with those at the same time on February 14. To facilitate this, they typically delved into the \textit{Outline View} to explore the events associated with the agent at these two distinct time points. Observing disparate agent behaviors across separate days, users inferred the existence of certain agent behavior patterns.  This discovery inspires us to further investigate strategies for visually ``folding'' the agent's timeline, such as overlaying two periods of the timeline, thereby aiding users in rapidly comparing and encapsulating the agent's behavior patterns.

\textbf{Enabling modifications for system configurations.}
Participants appreciated the aid provided by the novel behavior summarization method proposed in our study, which effectively mitigates information overload.  
Nevertheless, some users demonstrated an interest in understanding how these summaries are generated. They endorsed the summarization method after we clarified the details, as in \cref{fig:summary}. However, they still gave specific requirements, such as customizing the source of the summary contents. For example, one participant exhibited indifference towards the agent's location information. Such feedback motivates us to enable users to tailor the extraction pipeline in future research, thereby enhancing the usability of the exploratory analysis in a user-centric manner.

\subsection{Generalizability}

Our work builds upon the existing LLMAS, designed for the surveillance and analysis of agent behaviors. While we conduct our research based on Reverie, it can be seamlessly integrated into other LLMAS analysis processes. Moreover, the key components of our system, such as the \textit{Outline View} and \textit{Agent View}, are decoupled from the LLMAS implementations. 
The \textit{Monitor View} is a representation of the replay monitor ubiquitous in most LLMAS. Developers can easily provide their own monitoring snapshots to populate this view.
Therefore, our work is general to various LLMAS and can be used directly by developers in their LLMAS.

Our system’s capabilities extend beyond LLMAS analysis and can be applied to a wide range of applications, such as the analysis of multi-person communities and the development of open-world games.
For the analysis of multi-person communities, the \textit{Outline View} and \textit{Monitor View} can assist in simultaneously examining numerous actions on multiple subject timelines. This enables analysts to rapidly comprehend the main behaviors of different entities and their interactions.
Within the realm of open-world games, the incorporation of the \textit{Outline View} can aid players in exploring non-player characters (NPC) behaviors in an immersive manner. Game developers can also utilize the \textit{Agent View} to analyze and optimize the NPCs in the development stages, fostering the creation of more intelligent NPCs.

\subsection{Limitations and Future Work}

Despite the encouraging performance of \sysname{}, there are several limitations and potential areas for further research.

\textbf{Provide a more flexible interface.} The current layout of the agent line and position block in the \textit{Outline View} is pre-computed. 
Despite considerable efforts to minimize the crossover of lines, it remains difficult to avoid, particularly as the number of agents and the evolutionary timespan of LLMAS increase.
One of our future tasks is to provide a more flexible layout for the \textit{Outline View}, automatically reorganizing the view based on the user's interest regarding agent events.

\textbf{Allow users to modify pre-configured settings.} \sysname{} introduces a set of pre-configured settings for users, such as the granularity of \textit{Timeline Segmentation} and the similarity threshold for \textit{Cause Trace}. These configurations optimize the exploration experience for users, making better trade-offs between the intricate nature of the information and its succinct presentation. Nonetheless, some users expressed a desire to modify these presets during the analysis process to facilitate more flexible exploration. To accommodate these needs, we plan to incorporate a customizable preset panel for users in our system.

\note{\textbf{Support interactive exploration among different agent execution strategies.}
In this work, we focus on facilitating users' exploration and analysis of the LLMAS operational process. However, this process is significantly influenced by agent execution strategies like planning methods and memory mechanisms. For example, the agent may choose to first make a high-level plan to divide tasks into several sub-tasks that can be completed in different orders, or choose to adopt a depth-first strategy that adaptively changes its target based on the incoming information. While the design of an effective agent planning strategy is attracting an increasing amount of research attention \cite{chen2023agentverse,git2023xagent,git2023researchagent,zheng2023okragent}, how to interactively analyze the effect of different planning strategies in LLMAS is still unexplored. Moreover, analyzing the influence of agent memory mechanisms on the agent execution process is an area of considerable interest. While currently the agent memory mechanisms are usually hard-coded in the LLMAS program, allowing users to interactively modify the agent's memory content or recall strategies and visually examine its downstream effects could be crucial for better understanding and optimizing LLMAS.}

\note{
\textbf{Extend to multimodal LLMAS.}
Text-based interaction has been widely adopted in most existing LLMAS \cite{lin2023agentsims,park2023generative,qian2023communicative} in which agents are predicated on textual perception and decision-making.
Even embodied agents \cite{zhang2023building,zhu2023ghost} typically transmute the perceived multimodal data like imagery and auditory inputs into a textual format for later processing. 
However, with the popularity of multimodal LLMs\cite{openai2023gpt4,yang2023dawn}, the future may see the emergence of LLMAS in which agents genuinely perceive, think, and act based on multimodal data. 
Future work can explore how the agents interact with multimodal data (e.g., image interpretation \cite{feng_ipoet_2022} and creation \cite{feng2023promptmagician}) in this authentic multimodal LLMAS.
}
\section{Conclusion}
This work presents a visualization approach for LLMAS, addressing the challenge of analyzing complex agent behaviors during LLMAS evolution.
We introduce a general pipeline that establishes a hierarchical behavior structure from the raw execution events of LLMAS, including a behavior summarization algorithm and a cause-tracing method.
Our system, \sysname{}, offers an intuitive and hierarchical representation of the evolution of multiple agents, enabling users to interactively investigate behavior details and causes.
Through two usage scenarios and a user study, we have demonstrated the performance of our pipeline and visual designs.

\section*{Acknowledgments}
We would like to thank Ke Wang and Minfeng Zhu for their kind help.
We also would like to thank the anonymous reviewers for their insightful comments.
This paper is supported by the National Natural Science Foundation of China (62132017, 62302435), Zhejiang Provincial Natural Science Foundation of China (LD24F020011), and ``Pioneer'' and ``Leading Goose'' R\&D Program of Zhejiang (2024C01167).



\bibliographystyle{IEEEtran}
\bibliography{main}

\newpage

\section{Biography Section}
 

\begin{IEEEbiography}[{\includegraphics[width=1in,height=1.25in,clip,keepaspectratio]{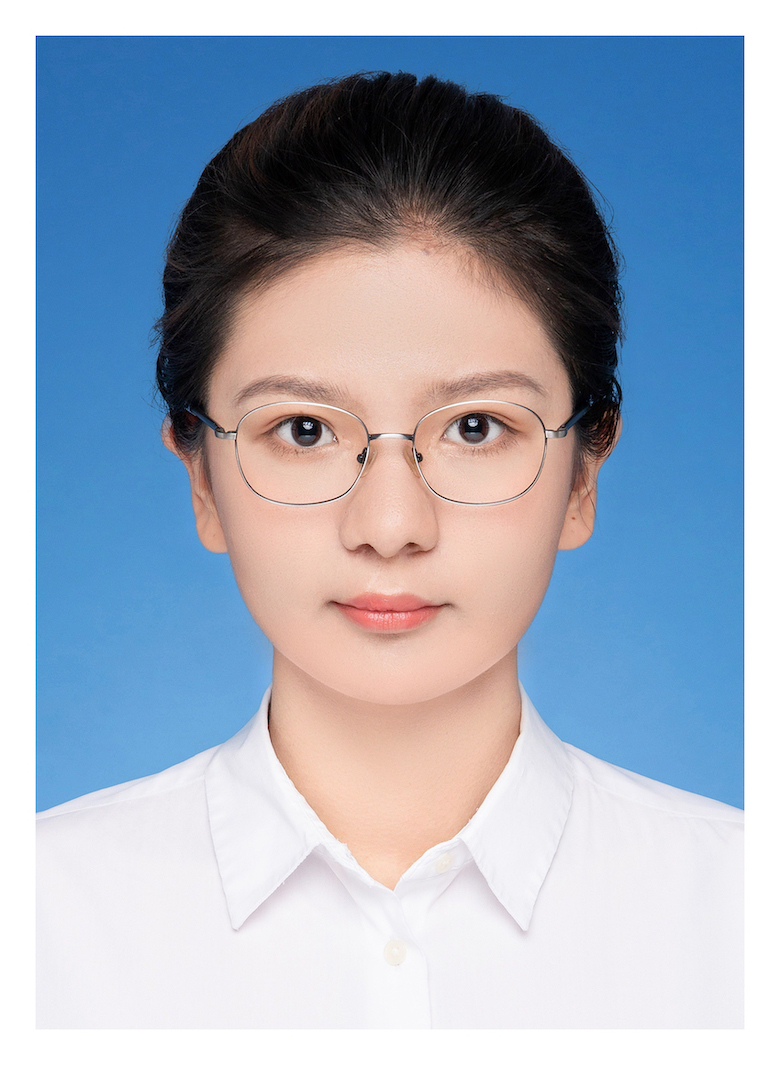}}]{Jiaying Lu}
is currently a Master student in the State Key Lab of CAD\&CG at Zhejiang University, China. She received the B.E. degree in Computer Science and Technology from the Zhejiang University, China in 2022. Her research interests include LLM agent and visual analytics.
\end{IEEEbiography}

\begin{IEEEbiography}[{\includegraphics[width=1in,height=1.25in,clip,keepaspectratio]{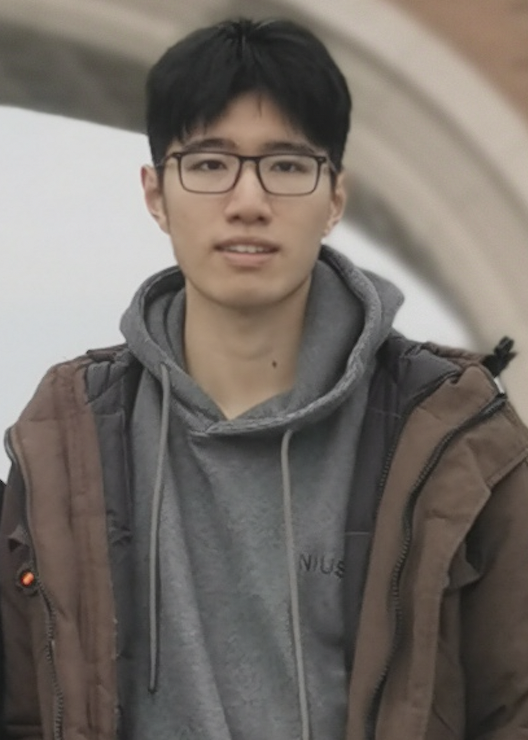}}]{Bo Pan} is currently a Ph.D. candidate in the State Key Lab of CAD\&CG at Zhejiang University, China. He received the BS degree in Electrical and Computer Engineering from the University of Illinois Urbana-Champaign and Zhejiang University in 2022. His research interests include visualization and deep learning.
\end{IEEEbiography}

\begin{IEEEbiography}[{\includegraphics[width=1in,height=1.25in,clip,keepaspectratio]{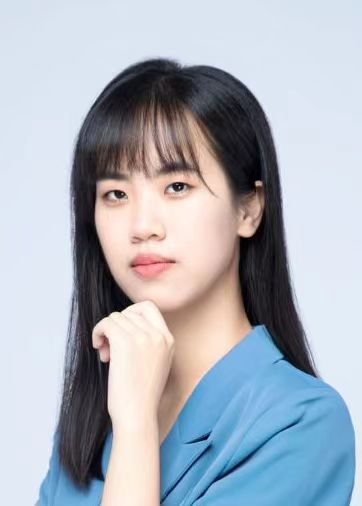}}]{Jieyi Chen}
is currently a Master student in the State Key Lab of CAD\&CG at Zhejiang University, China. She received the B.E. degree from the Zhejiang University of Technology, China in 2023. Her research interests include visualization and visual analytics.
\end{IEEEbiography}

\begin{IEEEbiography}[{\includegraphics[width=1in,height=1.25in,clip,keepaspectratio]{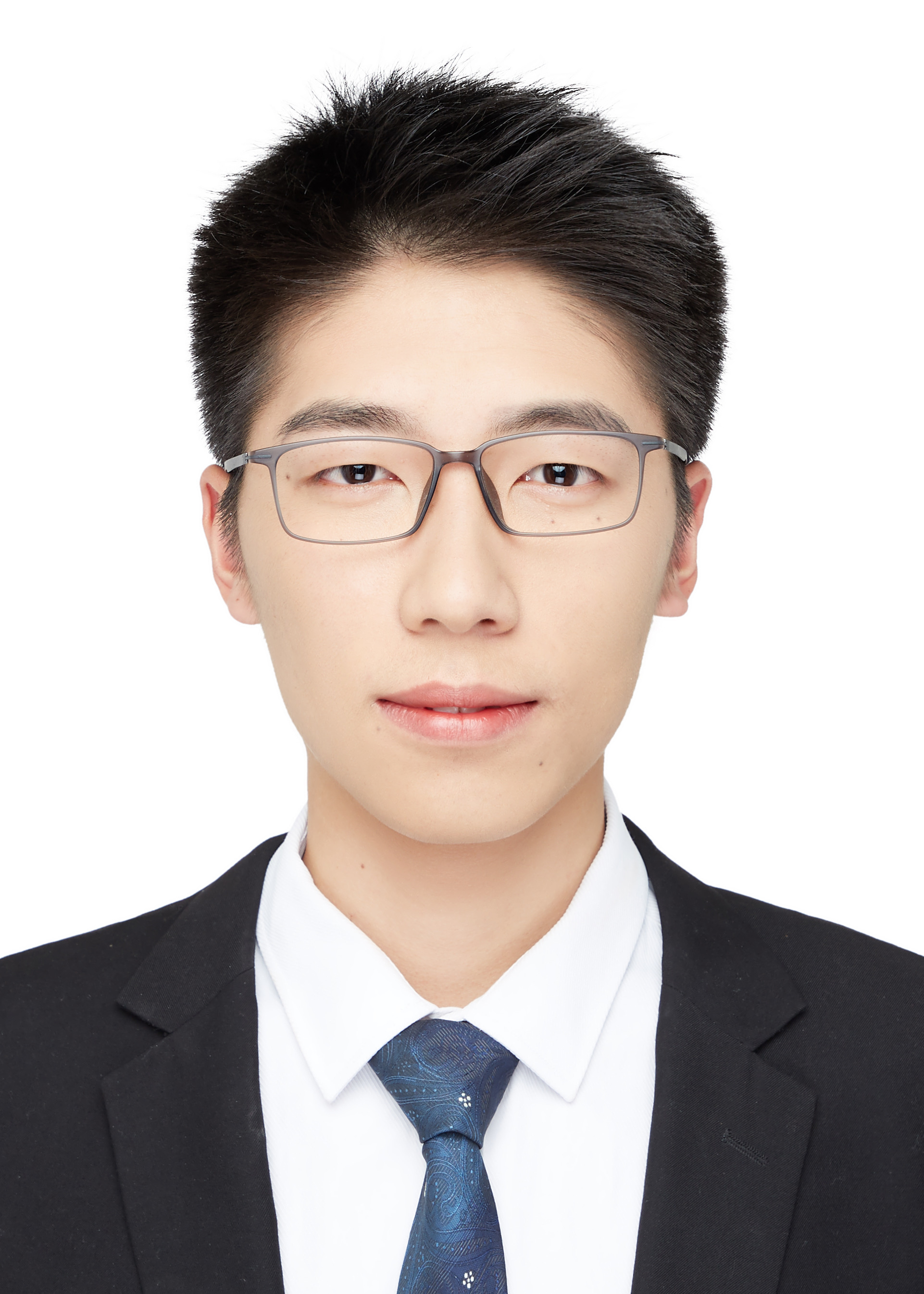}}]{Yingchaojie Feng} is currently a Ph.D. candidate in the State Key Lab of CAD\&CG at Zhejiang University, China. He received the B.E. degree in software engineering from the Zhejiang University of Technology, China in 2020. His research interests include data visualization, human-computer interaction, and natural language processing. For more details, please refer to https://yingchaojiefeng.github.io/.
\end{IEEEbiography}

\begin{IEEEbiography}[{\includegraphics[width=1in,height=1.25in,clip,keepaspectratio]{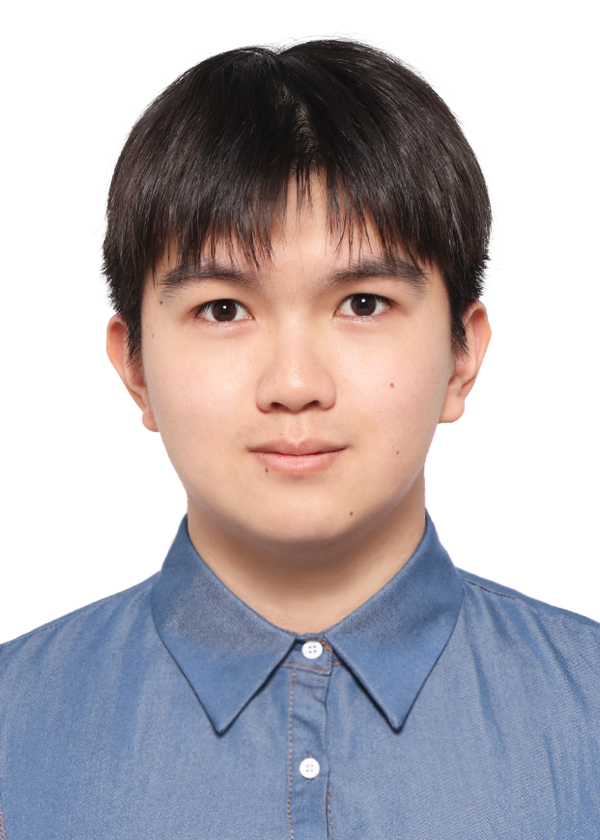}}]{Jingyuan Hu}
is an undergraduate in the Chu Kochen Honors College at Zhejiang University. His research interests include visualization and visual analytics.
\end{IEEEbiography}

\begin{IEEEbiography}[{\includegraphics[width=1in,height=1.25in,clip,keepaspectratio]{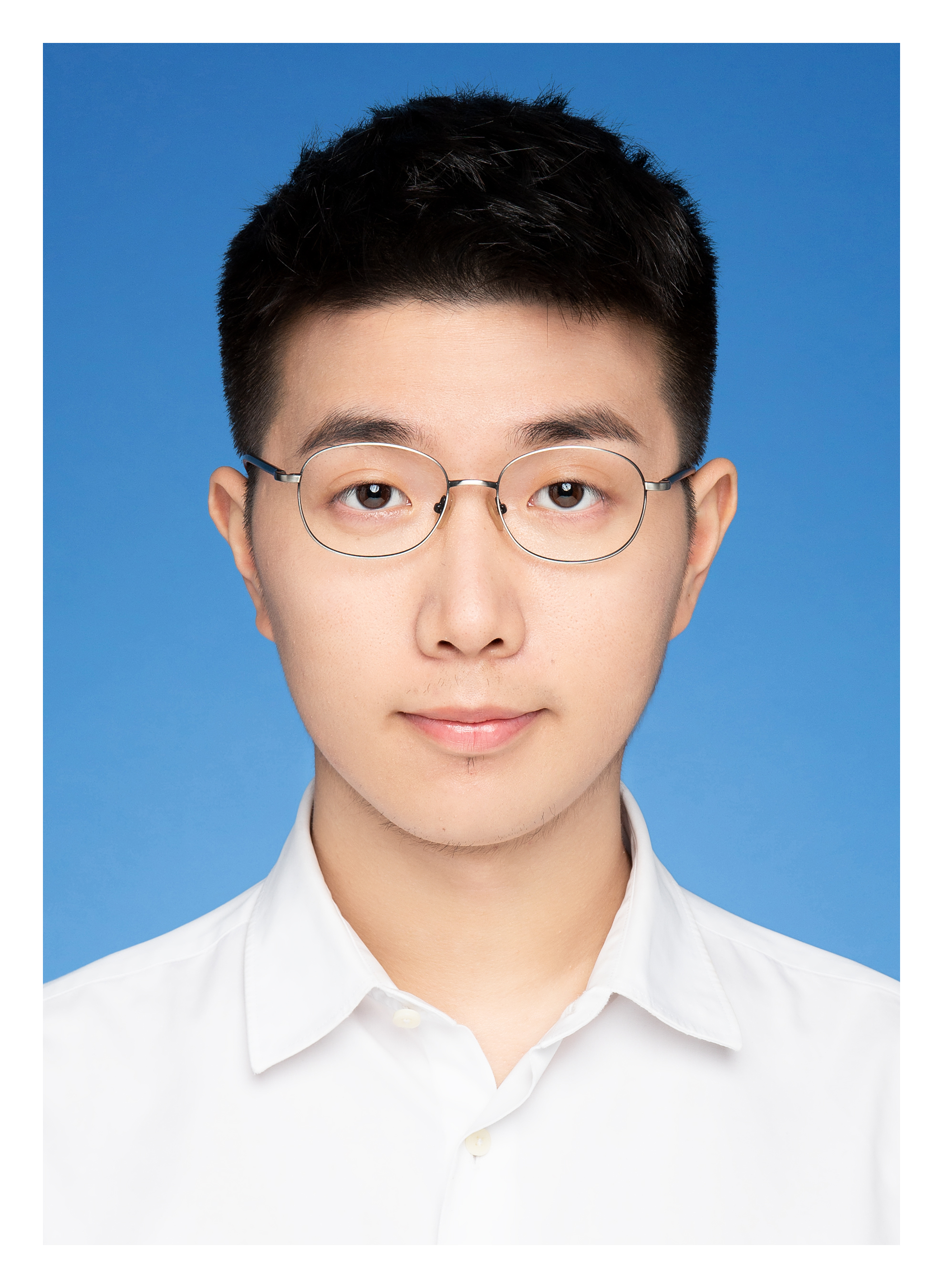}}]{Yuchen Peng} is currently a Ph.D candidate in the State Key Laboratory of Blockchain and Data Security from the Zhejiang University. He received the B.E. degree in computer science and technology from the Zhejiang University, China in 2022. His research interests include database system and data management in machine learning.
\end{IEEEbiography}

\begin{IEEEbiography}[{\includegraphics[width=1in,height=1.25in,clip,keepaspectratio]{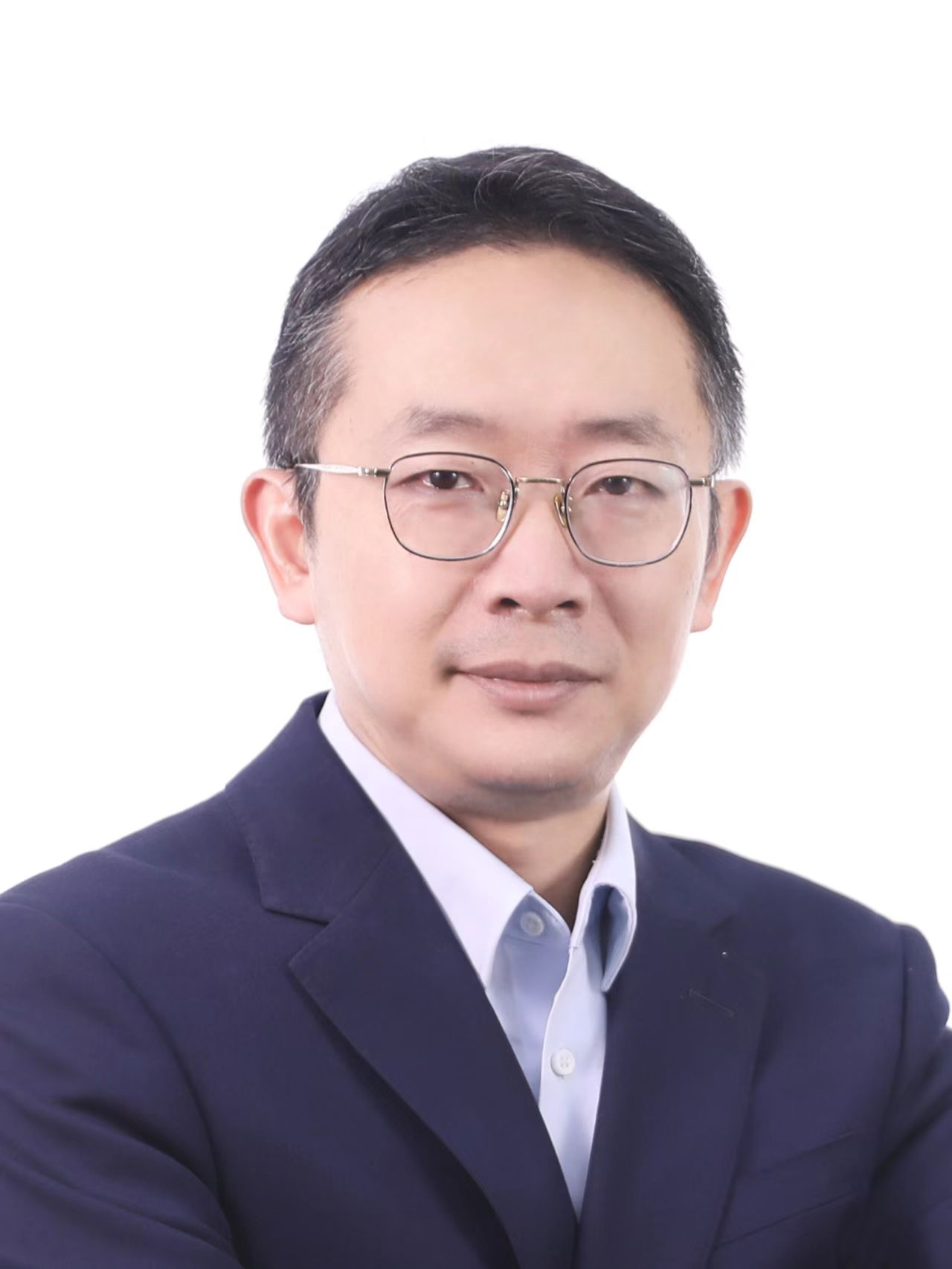}}]{Wei Chen}
is a professor in the State Key Lab of CAD\&CG at Zhejiang University. His current research interests include visualization and visual analytics. He has published more than 80 IEEE/ACM Transactions and IEEE VIS papers. He actively served in many leading conferences and journals, like IEEE PacificVIS steering committee, ChinaVIS steering committee, paper cochairs of IEEE VIS, IEEE PacificVIS, IEEE LDAV and ACM SIGGRAPH Asia VisSym. He is an associate editor of IEEE TVCG, IEEE TBG, ACM TIST, IEEE T-SMC-S, IEEE TIV, IEEE CG\&A, FCS, and JOV. More information can be found at: \url{http://www.cad.zju.edu.cn/home/chenwei}.
\end{IEEEbiography}

\vspace{11pt}


\vfill

\end{document}